\documentclass[prd,onecolumn,superscriptaddress,showpacs,nofootinbib,preprintnumbers]{revtex4}
\usepackage{amsmath}
\usepackage{amsfonts}
\usepackage{graphicx}
\usepackage{dcolumn}
\usepackage{hyperref}
\def\tr{\tilde{r}}

\def\rd{{\rm d}}

\def\be{\begin{equation}}
\def\ee{\end{equation}}
\def\bea{\begin{eqnarray}}
\def\eea{\end{eqnarray}}

\def\5{\overline 5}
\def\vp{\varphi}

\begin{document}

\title{Constraints on scalar-tensor models of dark energy \\
from observational and local gravity tests}

\author{Shinji Tsujikawa\footnote{shinji@nat.gunma-ct.ac.jp}}
\affiliation{Department of Physics, Gunma National College of
Technology, Gunma 371-8530, Japan}

\author{Kotub Uddin\footnote{k.uddin@qmul.ac.uk}}
\affiliation{School of Mathematical Sciences,\\
Queen Mary, University of London,
London E1 4NS, UK}

\author{Shuntaro Mizuno\footnote{mizuno@resceu.s.u-tokyo.ac.jp}}
\affiliation{Research Center for the Early Universe (RESCEU),
Graduate School of Science, \\
The University
of Tokyo, Tokyo 113-0033, Japan}

\author{Reza Tavakol\footnote{r.tavakol@qmul.ac.uk}}
\affiliation{School of Mathematical Sciences,\\
Queen Mary, University of London,
London E1 4NS, UK}

\author{Jun'ichi Yokoyama\footnote{yokoyama@resceu.s.u-tokyo.ac.jp}}
\affiliation{Research Center for the Early Universe (RESCEU),
Graduate School of Science, \\
The University
of Tokyo, Tokyo 113-0033, Japan}

\date{\today}

\vskip 1pc
\begin{abstract}

We construct a family of viable scalar-tensor models of dark energy (DE)
which possess a phase of late-time acceleration
preceded by a standard matter era, while at the
same time satisfying the local gravity constraints (LGC).
The coupling $Q$ between the scalar field
and the non-relativistic matter in the Einstein frame
is assumed to be constant in our scenario, which is
a generalization of $f(R)$ gravity theories
corresponding to the coupling $Q=-1/\sqrt{6}$.
We find that these models can be made compatible with local gravity
constraints even when $|Q|$ is of the order of unity through a chameleon
mechanism, if the scalar-field potential is chosen to have a
sufficiently large mass in the high-curvature regions.
We show that these models generally lead to
the divergence of the equation of state of DE,
which occur at smaller redshifts as the deviation from
the $\Lambda$CDM model become more significant.
We also study the evolution of matter density
perturbations and employ them to
place bounds on the coupling $|Q|$
as well as model parameters of the field potential from observations
of the matter power spectrum and the CMB anisotropies.
We find that, as long as $|Q|$ is smaller than
the order of unity, there exist allowed parameter
regions that are consistent with  both observational
and local gravity constraints.

\end{abstract}


\maketitle
\vskip 1pc

\section{Introduction}

The origin of dark energy (DE) has persistently posed
one of the most serious mysteries in modern
cosmology \cite{review,CST}.
The first step toward understanding the nature of DE
is to clarify whether it is a simple cosmological constant
or it originates from other sources that dynamically change
in time.  The dynamical DE models can be distinguished
from cosmological constant by studying the variation of
the equation of state of DE ($=w_{\rm DE}$) as well as
the evolution of density perturbations.
The scalar field models of DE such as quintessence \cite{quin}
and k-essence \cite{kes} predict a wide variety of variations
in $w_{\rm DE}$,  but still the current observational data
are not sufficient to rule out such models unless
the equation of state shows a peculiar evolution.
Moreover, the scalar field is required to have a light mass
$m_\phi$ comparable to the present Hubble parameter
($m_\phi \sim 10^{-33}$ eV), in order to give rise to an accelerated
expansion. This requirement is generally difficult to reconcile
with fifth-force experiments unless there exists some
mechanism by which the interaction range of
the scalar-field mediated force can be made shorter.

There exists another class of dynamical DE models that modify
Einstein gravity. The simplest models that belong to this
class are those that are based on the so called $f(R)$
gravity theories in which the Lagrangian density $f$ 
is a function of the Ricci scalar $R$.
It is well known that theories of the type $f(R)=R+\alpha R^2$
can give rise to an inflationary expansion in the early universe
because of the dominance of the $\alpha R^2$ term \cite{Star80}.
In the context of DE, the model $f(R)=R-\mu^{2(n+1)}/R^n$
($n>0$) was proposed to explain the late-time accelerated
expansion due to the dominance of
the term  $\mu^{2(n+1)}/R^n$ \cite{Capo}
(see also Refs.~\cite{fRearly}).
It was found, however, that this model is plagued by
a number of problems such as the instability
of matter perturbations \cite{Dolgov} as well as the absence of
a matter-dominated epoch \cite{APT}.

In the past few years there has been a burst of activity
in the search for viable $f(R)$ DE
models \cite{fRpapers,AGPT,Max,Song,linear,Star07,Hu07,Appleby,Tsuji08,TUT}.
In Ref.~\cite{AGPT} the conditions for the cosmological
viabilities of $f(R)$ DE models (having a matter era
followed by an accelerated epoch) were derived
without specifying the form of $f(R)$.
A number of general conditions are required
on general viability and stability grounds.
For the existence of a prolonged saddle matter era,
the quantity $m=Rf_{,RR}/f_{,R}$
needs to be positive and close to 0.
To avoid anti-gravity, $f_{,R}$ is required to be positive in regions
$R \ge R_1$, where $R_1~(>0)$ is a Ricci scalar
at a de-Sitter attractor responsible for the accelerated
expansion. Also, to ensure that
density perturbations do not exhibit violent
instabilities, we require $f_{,RR}>0$ \cite{Song,linear}.
The conditions $f_{,RR}>0$ and $f_{,R}>0$
(for $R>R_1$) have been shown to also ensure the absence of
ghosts and tachyons \cite{Star07}.

The local gravity constraints (LGC) should also be
satisfied for the viability of $f(R)$ models \cite{lgcpapers}.
The $f(R)$ gravity in the metric formalism
is equivalent to scalar-tensor theory with no scalar
kinetic term, namely, the Brans-Dicke model
with a potential and $\omega_{\rm BD}=0$ \cite{Chiba}.
If the mass of the scalar-field degree of freedom
always remains as light as the present Hubble parameter $H_0$,
one can not satisfy the LGC due to the appearance of the
long-ranged fifth force.
It is possible to design the field potential so that
the mass of the field is heavy in a large-curvature region
where local gravity experiments are carried out.
Then the interaction range of the fifth force
becomes short in such a high-density
region, which allows the possibility of the models
being compatible with LGC.

In fact a number of viable models based on $f(R)$ theories
have been proposed \cite{Hu07,Star07,Appleby,Tsuji08}
that can satisfy both the cosmological and local gravity constraints
discussed above. In the high-density region ($R \gg R_c$)
these models have asymptotic behavior
$f(R) \simeq R-\mu R_c \left[1-(R/R_c)^{-2n}
\right]$ ($\mu>0$, $R_c>0$, $n>0$),
where $R_c$ is of the order of the present Ricci scalar.
Inside a spherically symmetric body with an energy density $\rho_m$,
the field acquires a minimum at $R \simeq \rho_m$ with a mass
much heavier than $H_0$.
In this case the body has a thin-shell inside it so that
the effective coupling between the field and the matter
decreases through the so-called
chameleon mechanism \cite{KW1,KW2,Nava,Max,CT}.
The bounds on the model parameters of such models derived from
solar-system and equivalence principle constraints are given by
$n>0.5$ and $n>0.9$, respectively \cite{CT}.
For viable $f(R)$ models there are also a number of
interesting observational signatures
such as the divergence of the equation of state of DE \cite{AT,Tsuji08}
and the peculiar evolution of matter
perturbations \cite{Hu07,Star07,Tsuji08,TUT}.
This is useful to distinguish $f(R)$ gravity models
from the $\Lambda$CDM model.

In the Einstein frame the $f(R)$ gravity corresponds to a constant
coupling $Q=-1/\sqrt{6}$ between dark energy and
the non-relativistic fluid \cite{APT} (see Eq.~(\ref{Q}) for
the definition of $Q$).
Basically, this is equivalent to the coupled quintessence
scenario \cite{lucacoupled} with a specific coupling.
Our aim in this paper is to generalize the analysis
to scalar-tensor theories with the action (\ref{action2})
in which the coupling $Q$ is an arbitrary constant.
We regard the Jordan frame as a physical one in which
the usual matter conservation law holds.
The dark energy dynamics in scalar-tensor theories has been
investigated in many papers \cite{DEeos,stpapers,Jerome,GPRS,Peri}
after the pioneering works of Refs.~\cite{stori,Boi}.
If the mass of the field $\phi$ is always of the order of $H_0$,
the solar-system constraint
$\omega_{\rm BD}>4.0 \times 10^4$ \cite{hoyle}
gives the bound $|Q|<2.5 \times 10^{-3}$.
Previous studies dealing with the compatibility
of the scalar-tensor DE models with LGC have
restricted their analysis to this small coupling region \cite{Jerome,GPRS}.
We wish to extend the analysis to the case in which
the coupling $|Q|$ is larger than the above massless bound.
In fact one can design the potential $V(\phi)$ so that
the mass of the field is sufficiently heavy in the high-density region
to satisfy LGC through the chameleon mechanism.
We shall construct such a viable field potential inspired by
the case of the $f(R)$ gravity and place experimental
bounds on model parameters in terms of the function of $Q$.

We shall also study the variation of the equation of state for DE
and the evolution of density
perturbations in such scalar-tensor theories.
Interestingly, we find that the divergent behavior of $w_{\rm DE}$
is also present as in the case of $f(R)$ gravity.
We also estimate the growth rate of matter perturbations 
and show that the non-standard
evolution of perturbations manifests itself from a certain
epoch (depending upon model parameters) during the matter era.
This is useful to place constraints on model parameters
using the data of large scale structure and CMB.

This paper is organized as follows.
In Sec.~\ref{secmodel} we consider a class of scalar-tensor theories
with constant coupling $Q$.
In Sec.~\ref{sechomcos} we study the background cosmological
dynamics and consider the case of constant as well as varying $\lambda$
(the slope of the potential in the physical frame).
In this section we also introduce a family of potentials
which are natural generalizations of a viable family of models
in $f(R)$ gravity.
In Sec.~\ref{secLGC} we discuss the LGC under
the chameleon mechanism and place experimental bounds
on parameters of viable scalar-tensor models
using solar-system and equivalence principle constraints.
In Sec.~\ref{secSN} we study the evolution of the equation of state of DE
and show that the divergence of $w_{\rm DE}$ previously
found in $f(R)$ theories is also  present
in the class of scalar-tensor models considered here which are
compatible with LGC.
In Sec.~\ref{secmatter} we discuss the evolution of
density perturbations and place constraints on the coupling and
model parameters employing the predicted
difference in the slopes of the power spectra between
large scale structure and the CMB.
Finally we conclude in Sec.~\ref{conclude}.
The stability analysis
which is crucial to derive the background
cosmological scenario is briefly summarized in
Appendix~\ref{apstab_ana}.
In Appendix~\ref{des_st} we also clarify the stability
condition of a de-Sitter point that appears
in the presence of the coupling $Q$ for varying $\lambda$.

\section{Scalar-tensor theories}
\label{secmodel}

We start with a class of scalar-tensor theories,
which includes the pure $f(R)$ theories as well as 
the quintessence models as special cases, in the form
\begin{eqnarray}
\label{action}
S =  \int {\rm d}^4 x\sqrt{-g} \left[ \frac12 f(\vp, R)
-\frac12 \zeta (\vp) (\nabla \vp)^2 \right]
+S_m (g_{\mu \nu}, \Psi_m)\,.
\end{eqnarray}
Here, $f$ is a general differentiable function of the scalar field
$\vp$ and the Ricci scalar $R$, $\zeta$ is a differentiable function of
$\vp$ and $S_m$ is a matter Lagrangian that depends on the
metric $g_{\mu \nu}$ and matter fields $\Psi_m$.
We also choose units such that $\kappa^2 \equiv
8\pi G=1$, and restore the gravitational constant $G$ when
it makes the discussion more transparent.

The action (\ref{action}) can be transformed to the
so called Einstein frame under the conformal transformation \cite{Maeda89}
:
\begin{eqnarray}
\tilde{g}_{\mu \nu}=e^{2\Omega}\,g_{\mu \nu}\,,
\end{eqnarray}
where
\begin{eqnarray}
\Omega=\frac12\,{\rm ln}\,F,\qquad
F=\frac{\partial f}{\partial R}\,.
\end{eqnarray}
In the following we shall consider $F$ to be positive
in order to ensure that gravity is attractive.

We shall be considering theories of the type
\begin{eqnarray}
\label{stensor}
f(\vp, R)=F(\vp)R-2V(\vp)\,,
\end{eqnarray}
for which the conformal factor $\Omega$ depends
upon $\vp$ only.
Introducing a new scalar field $\phi$ by
\begin{eqnarray}
\label{phire}
\phi=\int \left [\sqrt{\frac32 \left(\frac{F_{,\vp}}{F}\right)^2+ \frac{\zeta}{F}}\, \right]
{\rm d}\vp\,,
\end{eqnarray}
the action in the Einstein frame becomes \cite{Maeda89}
\begin{eqnarray}
\label{SE}
S_E=\int {\rm d}^4 x \sqrt{-\tilde{g}}
\left[ \frac12 \tilde{R}-\frac12
(\tilde{\nabla}\phi)^2
-U(\phi) \right]+S_m (\tilde{g}_{\mu \nu}F^{-1}, \Psi_m)\,,
\end{eqnarray}
where a tilde represents quantities in the
Einstein frame and
\begin{eqnarray}
U=\frac{V}{F^2}\,.
\end{eqnarray}

In $f(R)$ gravity theories without the field $\vp$,
the conformal factor $\Omega$ depends only on $R$.
Introducing a new scalar field to be
\begin{eqnarray}
\label{fRre}
\phi=\frac{\sqrt{6}}{2}\,{\rm ln}\,F\,,
\end{eqnarray}
the action in the Einstein frame is given by (\ref{SE}) with
the potential \cite{Maeda89}
\begin{eqnarray}
U=\frac{RF-f}{2F^2}\,.
\end{eqnarray}
Hence, the $f(R)$ gravity can be cast in the
form of scalar-tensor theories of the type
(\ref{action}) with (\ref{stensor}), by identifying
the potential in the Jordan frame to be
$V=(RF-f)/2$.

In order to describe the strength of the coupling
between dark energy and a non-relativistic matter,
we introduce the following quantity
\begin{eqnarray}
\label{Q}
Q=-\frac{F_{,\phi}}{2F}\,.
\end{eqnarray}
{} From Eq.~(\ref{fRre}) one has $F=e^{2\phi/\sqrt{6}}$
which shows that the $f(R)$ gravity corresponds to
\begin{eqnarray}
Q=-1/\sqrt{6}\,.
\end{eqnarray}
In what follows we shall study a class of scalar-tensor theories
where $Q$ is treated as an arbitrary constant. This class includes
a wider family of models, including $f(R)$ gravity,
induced gravity and quintessence models.
Using Eqs.~(\ref{phire}) and (\ref{Q}) we have the following relations
\begin{eqnarray}
F=e^{-2Q \phi}\,, \quad
\zeta=(1-6Q^2)F \left( \frac{{\rm d} \phi}
{{\rm d}\vp} \right)^2\,.
\label{conf_factor}
\end{eqnarray}
Then action (\ref{action}) in the Jordan frame together with
(\ref{stensor}) yields
\begin{eqnarray}
\label{action2}
S =  \int {\rm d}^4 x\sqrt{-g} \Bigg[ \frac12 F R
-\frac12 (1-6Q^2)F(\nabla \phi)^2 -V \Bigg]
+S_m (g_{\mu \nu}, \Psi_m)\,.
\end{eqnarray}
In $f(R)$ gravity the kinetic term of the field $\phi$ vanishes
with the potential given by $V=(RF-f)/2$.
Note that in the limit, $Q \to 0$, the action (\ref{action2})
reduces to the one for a minimally coupled scalar field $\phi$
with a potential $V(\phi)$.

It is informative to compare (\ref{action2}) with the following action
\begin{eqnarray}
\label{action3}
S =  \int {\rm d}^4 x\sqrt{-g} \Bigg[ \frac12 \chi R
-\frac{\omega_{\rm BD}}{2\chi} (\nabla \chi)^2 -V \Bigg]
+S_m (g_{\mu \nu}, \Psi_m)\,,
\end{eqnarray}
which corresponds to Brans-Dicke theory
with a potential $V$.
Setting $\chi=F=e^{-2Q\phi}$, one
easily finds that two actions are equivalent if the
parameter $\omega_{\rm BD}$ is related with
$Q$ via the relation
\begin{eqnarray}
\label{BD}
3+2\omega_{\rm BD}=\frac{1}{2Q^2}\,.
\end{eqnarray}
Under this condition, the theories given by (\ref{action2})
are equivalent to the Brans-Dicke theory with a potential $V$.

In the following sections we shall in turn consider
the evolution of the background dynamics in homogeneous settings,
the local gravity constraints and
the matter density perturbations.

\section{Homogeneous cosmology}
\label{sechomcos}

In what follows we shall discuss cosmological dynamics
for the action (\ref{action2}) in the flat
Friedmann-Lemaitre-Robertson-Walker (FLRW) spacetime
${\rm d}s^2=-{\rm d} t^2+a^2(t){\rm d}{\bf x}^2$,
where $t$ is cosmic time and $a(t)$ is the scale factor.
As a source of the matter action $S_{m}$, we consider
a non-relativistic fluid with energy density $\rho_{m}$
and a radiation with energy density $\rho_{\rm rad}$.
Then the evolution equations in the Jordan frame
are given by
\begin{eqnarray}
\label{be1}
& & 3FH^2=\frac12 (1-6Q^2) F\dot{\phi}^2+V
-3H\dot{F}+\rho_{m}+\rho_{\rm rad}\,, \\
\label{be2}
& & 2F\dot{H}=-(1-6Q^2)F \dot{\phi}^2
-\ddot{F}+H\dot{F}-\rho_{m}
-\frac43 \rho_{\rm rad}\,, \\
\label{be3}
& & \dot{\rho}_{m}+3H \rho_{m}=0\,,\\
\label{be4}
& & \dot{\rho}_{\rm rad}+4H \rho_{\rm rad}=0\,,
\end{eqnarray}
where $H\equiv\dot{a}/a$ and a dot represents
a derivative with respect to $t$.

Taking the time-derivative of Eq.~(\ref{be1})
and using Eq.~(\ref{be2}), we obtain
\begin{eqnarray}
\label{be5}
(1-6Q^2) F \left( \ddot{\phi}
+3H\dot{\phi}+\frac{\dot{F}}{2F}\dot{\phi}
\right)+V_{,\phi}+QFR=0\,,
\end{eqnarray}
where the Ricci scalar is given by
\begin{eqnarray}
\label{Ricci}
R=6(2H^2+\dot{H})\,.
\end{eqnarray}
We regard the Jordan frame as a physical one,
since the usual matter conservation
holds in this frame [see Eq.~(\ref{be3})].

In order to study the cosmological dynamics, it is
convenient to introduce the
following dimensionless phase space variables
\begin{eqnarray}
x_1=\frac{\dot{\phi}}{\sqrt{6}H}\,, \quad
x_2=\frac{1}{H}\sqrt{\frac{V}{3F}}\,, \quad
x_3=\frac{1}{H}\sqrt{\frac{\rho_{\rm rad}}{3F}}\,.
\end{eqnarray}
Then the constraint equation (\ref{be1}) yields
\begin{eqnarray}
\label{con}
\Omega_{m} \equiv \frac{\rho_{m}}{3FH^2}=
1-(1-6Q^2)x_1^2-x_2^2-2\sqrt{6}Q x_1-x_3^2\,.
\end{eqnarray}
We also define the following quantities
\begin{eqnarray}
\Omega_{\rm rad} \equiv x_3^2\,,\quad
\Omega_{\rm DE} \equiv (1-6Q^2)x_1^2+x_2^2+
2\sqrt{6}Q x_1\,.
\end{eqnarray}
Eq.~(\ref{con}) then yields the relation 
$\Omega_m+\Omega_{\rm rad}+\Omega_{\rm DE}=1$.

{} From Eqs.~(\ref{be2}) and (\ref{be5}) we obtain
\begin{eqnarray}
\label{be6}
\frac{\dot{H}}{H^2}&=& -\frac{1-6Q^2}{2} \left[
3+3x_1^2-3x_2^2+x_3^2-6Q^2 x_1^2
+2\sqrt{6}Q x_1 \right]+
3Q (\lambda x_2^2-4Q)\,, \\
\frac{\ddot{\phi}}{H^2} &=&
3(\lambda x_2^2-\sqrt{6} x_1)+3Q
 \left[ (5-6Q^2)x_1^2+2\sqrt{6} Q x_1
 -3x_2^2+x_3^2-1 \right]\,.
\end{eqnarray}

Using these relations, we obtain the following
autonomous equations:
\begin{eqnarray}
\label{au1}
\frac{{\rm d} x_1}{{\rm d} N}
&=& \frac{\sqrt{6}}{2} (\lambda x_2^2-\sqrt{6} x_1)
+\frac{\sqrt{6}Q}{2} \left[ (5-6Q^2) x_1^2+
2 \sqrt{6}Q x_1-3x_2^2+x_3^2-1 \right]
-x_1 \frac{\dot{H}}{H^2}  \,, \\
\label{au2}
\frac{{\rm d} x_2}{{\rm d} N}
&=& \frac{\sqrt{6}}{2} (2Q-\lambda)x_1 x_2
-x_2 \frac{\dot{H}}{H^2} \,, \\
\label{au3}
\frac{{\rm d} x_3}{{\rm d} N}
&=& \sqrt{6} Q x_1 x_3-2x_3
-x_3 \frac{\dot{H}}{H^2} \,,
\end{eqnarray}
where $N\equiv{\rm ln}\,(a)$ is the number of e-foldings
and $\lambda$ is defined by

\begin{eqnarray}
\lambda\equiv-\frac{V_{,\phi}}{V}\,.
\end{eqnarray}
The exponential potential $V(\phi)=V_0 e^{-\lambda \phi}$
gives a constant value of $\lambda$.
Generally, however, $\lambda$ is dependent on $\phi$, where the field
$\phi$ is a function of $x_1$, $x_2$ and $x_3$ through
the definition of $x_2$ and Eq.~(\ref{be6}).
Hence Eqs.~(\ref{au1})-(\ref{au3}) are closed.
The effective equation of state is given by
\begin{eqnarray}
\label{weff}
w_{\rm eff} &\equiv&-1-\frac23 \frac{\dot{H}}{H^2}
\nonumber \\
&=& -1+\frac{1-6Q^2}{3}
(3+3x_1^2-3x_2^2+x_3^2-6Q^2 x_1^2+
2\sqrt{6} Qx_1 )-2Q (\lambda x_2^2-4Q)\,.
\end{eqnarray}

In what follows we shall first discuss the case of constant $\lambda$
and then proceed to consider the varying $\lambda$ case.

\subsection{Constant $\lambda$}

If $\lambda$ is a constant, one can derive the
fixed points of the system by setting the r.h.s.
of Eqs.~(\ref{au1})-(\ref{au3})
to be zero.
In the absence of radiation ($x_3=0$),
we obtain the following fixed points:
\begin{itemize}
\item (a) $\phi$ matter-dominated era ($\phi$MDE \cite{lucacoupled})
\begin{eqnarray}
\label{fp1}
(x_1,x_2)=\left( \frac{\sqrt{6}Q}{3(2Q^2-1)}, 0 \right)\,,
\quad \Omega_{m}=\frac{3-2Q^2}{3(1-2Q^2)^2}\,,
\quad w_{\rm eff}=\frac{4Q^2}{3(1-2Q^2)}\,.
\end{eqnarray}
\item (b1) Kinetic point 1
\begin{eqnarray}
(x_1,x_2)=\left( \frac{1}{\sqrt{6}Q+1}, 0 \right)\,,
\quad \Omega_{m}=0\,,
\quad w_{\rm eff}=\frac{3-\sqrt{6}Q}{3(1+\sqrt{6}Q)}\,.
\end{eqnarray}
\item (b2) Kinetic point 2
\begin{eqnarray}
(x_1,x_2)=\left( \frac{1}{\sqrt{6}Q-1}, 0 \right)\,,
\quad \Omega_{m}=0\,,
\quad w_{\rm eff}=\frac{3+\sqrt{6}Q}{3(1-\sqrt{6}Q)}\,.
\end{eqnarray}
\item (c) Scalar-field dominated point
\begin{eqnarray}
(x_1,x_2)=\left( \frac{\sqrt{6}(4Q-\lambda)}
{6(4Q^2-Q\lambda-1)}, \left[ \frac{6-\lambda^2+8Q\lambda
-16Q^2}{6(4Q^2-Q\lambda-1)^2} \right]^{1/2} \right)\,,
~~\Omega_{m}=0\,,
\quad w_{\rm eff}=-\frac{20Q^2-9Q\lambda-3+\lambda^2}
{3(4Q^2-Q\lambda -1)}\,.
\end{eqnarray}
\item (d) Scaling solution
\begin{eqnarray}
\label{scaling}
(x_1,x_2)=\left( \frac{\sqrt{6}}{2\lambda},
\left[ \frac{3+2Q\lambda -6Q^2}{2\lambda^2} \right]^{1/2}
\right)\,,
\quad \Omega_{m}=1-\frac{3-12Q^2+7Q\lambda}
{\lambda^2}\,,
\quad w_{\rm eff}=-\frac{2Q}{\lambda}\,.
\end{eqnarray}
\item (e) de-Sitter point (present for $\lambda=4Q$)
\begin{eqnarray}
\label{fp5}
(x_1,x_2)=(0,1)\,,
\quad \Omega_{m}=0\,,
\quad w_{\rm eff}=-1\,.
\end{eqnarray}
\end{itemize}
Note that, when $x_3 \neq 0$ we have a radiation
fixed point $(x_1,x_2,x_3)=(0,0,1)$.

One can easily confirm that the de-Sitter point exists for
$\lambda=4Q$, by setting $\dot{\phi}=0$
in Eqs.~(\ref{be1}), (\ref{be2}) and (\ref{be5}).
This de-Sitter solution appears in the presence of
the coupling $Q$.
Note that this is the special case of the scalar-field
dominated point (c).

Now given a value for $\lambda$, and using the
stability conditions for the above fixed
points given in the Appendix~A,
the cosmological dynamics can be specified.
We shall briefly discuss the cases $Q=0$ and $Q \neq 0$
in turn.

\subsubsection{$Q=0$}

When $Q=0$ (i.e., $F=1$, which corresponds to
a standard minimally coupled scalar field),
the eigenvalues $\mu_1$ and $\mu_2$ of the
Jacobian matrix for perturbations about the fixed points
reduce to those derived in Ref.~\cite{CLW} with $\gamma=1$
(see Ref.~\cite{Halli} for earlier works).
In this case the matter-dominated era corresponds to
either the point (a) or (d).
The point (a) is a saddle node because $\mu_1=-3/2$
and $\mu_2=3/2$.
The point (d) is stable for $\lambda^2>3$, in which
case $\Omega_{m}<1$.
The late-time accelerated expansion
($w_{\rm eff}<-1/3$) can be realized by using
the point (c), whose condition is given by $\lambda^2<2$.
Under this condition the point (c) is a stable node.
Hence, if $\lambda^2<2$,
the saddle matter solution (a) is followed by the
stable accelerated solution (c) [note that
in this case $\Omega_m<0$ for the point (d)].
The scaling solution (d) can have a matter era for $\lambda^2 \gg 1$,
but in this case the epoch following the matter era
is not of an accelerated nature.

\subsubsection{$Q \neq 0$}

We next consider the case of non-zero values of $Q$.
Here we do not consider the special case of $\lambda=4Q$.
If the point (a) is responsible for the matter-dominated
epoch, we require the condition $Q^2 \ll 1$.
We then have $\Omega_{m} \simeq 1+10Q^2/3>1$ and
$w_{\rm eff} \simeq 4Q^2/3$, for the $\phi$MDE.
When $Q^2 \ll 1$ the scalar-field dominated point (c)
yields an accelerated expansion provided that
$-\sqrt{2}+4Q<\lambda<\sqrt{2}+4Q$\footnote{Note that
under the condition $Q^2\ll 1$ and in the case
where the dynamics is in the accelerated epoch, the condition
$|Q\lambda|<1$ is also satisfied.}.
Under these conditions the $\phi$MDE point
is followed by the late-time acceleration.
It is worth noting that in the case of $f(R)$ gravity
($Q=-1/\sqrt{6}$) the $\phi$MDE point
corresponds to, $\Omega_m=2$ and $w_{\rm eff}=1/3$.
In this case the universe in the matter era prior to
late-time acceleration evolves as $a \propto t^{1/2}$,
which is different from the evolution in
the standard matter dominated epoch \cite{APT}.

We note that the scaling solution (d) can
give rise to the equation of state, $w_{\rm eff} \simeq 0$
for $|Q| \ll |\lambda|$.
In this case, however, the condition $w_{\rm eff}<-1/3$
for the point (c) gives $\lambda^2 \lesssim 2$.
Then the energy fraction of the pressureless matter
for the point (d) does not satisfy the condition
$\Omega_{m} \simeq 1$.
In summary the viable cosmological trajectory corresponds to the
sequence from the $\phi$MDE to the scalar-field
dominated point (c) under the conditions,
$Q^2 \ll 1$ and $-\sqrt{2}+4Q<\lambda<\sqrt{2}+4Q$.

\subsection{Varying $\lambda$}

When the time-scale of the variation of $\lambda$ is smaller than
that of the cosmic expansion, the fixed points derived above
in the case of constant  $\lambda$ can be regarded as
the ``instantaneous'' fixed points \cite{Maco}.
We shall briefly consider the cases of
$Q=0$ and $Q\neq 0$ in turn.

\subsubsection{$Q=0$}

We begin with a brief discussion of the $Q=0$ case.
If the condition $\lambda^2<2$ is satisfied
throughout the cosmic evolution, the cosmological
trajectory is similar to the constant $\lambda$ case discussed above
except for the fact that the fixed points are regarded as the
``instantaneous'' ones.
In this case the saddle matter solution (a)
is followed by the accelerated point (c).

When $\lambda^2 \gg 1$ the scaling solution (d) is
stable with $\Omega_{m} \simeq 1$.
Hence the cosmological trajectory during the matter era
chooses the scaling solution (d) rather than the saddle point (a).
If $|\lambda|$ decreases at late-times, such that it satisfies
the acceleration condition $\lambda^2<2$,
the trajectory stops following the solution represented by
the matter point (d) to
follow the scalar-field dominated point (c)\footnote{Note that
the de-Sitter solution (e) exists only for $\lambda=0$,
i.e., for the case of cosmological constant ($V=$const.).}.
A representative model of this type is provided
by the double exponential potential,
$V(\phi)=V_0 (e^{-\lambda_1 \phi}+e^{-\lambda_2 \phi})$,
with, $\lambda_1^2 \gg 1$ and $\lambda_2^2<2$ \cite{Nelson}.
The assisted quintessence models in Ref.~\cite{Kim}
also lead to a similar cosmological evolution.

\subsubsection{$Q \neq 0$}

We shall now proceed to consider the case of non-zero $Q$.
If $|\lambda|$ is initially much larger than unity and decreases
with time, it happens that the solutions finally approach
the de-Sitter solution (e) with $\lambda=4Q$.
As we shall show in the Appendix~\ref{des_st},
the de-Sitter point (e) is in fact stable even for
the variable $\lambda$ case, if the potential
satisfies the condition $Q ({\rm d}\lambda/{\rm d} F) > 0$ or
${\rm d} \lambda / {\rm d} \phi < 0$
at the de-Sitter point.

In the context of $f(R)$ gravity, it has been shown that the model,
\begin{eqnarray}
\label{fRmodel}
f(R)=R-\mu R_c\left[1-(R/R_c)^{-2n} \right] \quad
(\mu>0, R_c>0, n>0)\,,
\end{eqnarray}
is a good example which can be consistent
with cosmological and local gravity
constraints \cite{Tsuji08}.
Note that the models proposed by Hu \& Sawicki \cite{Hu07}
and Starobinsky \cite{Star07} reduce to this form of $f(R)$
in the high-curvature region ($R \gg R_c$).
In this model the field $\phi$ is related to the Ricci scalar $R$
via the relation $e^{2\phi/\sqrt{6}}=1-2n\mu (R/R_c)^{-(2n+1)}$.
Hence, the potential $V=(FR-f)/2$ can be expressed
in terms of the field $\phi$ as
\begin{eqnarray}
\label{fRpo}
V(\phi)=\frac{\mu R_c}{2}
\left[ 1-\frac{2n+1}{(2n \mu)^{2n/(2n+1)}}
\left( 1-e^{2\phi/\sqrt{6}} \right)^{2n/(2n+1)}
\right]\,.
\end{eqnarray}
The parameter $\lambda$ is then given by
\begin{eqnarray}
\lambda=-\frac{4n}{\sqrt{6}(2n \mu)^{2n/(2n+1)}}
e^{2\phi/\sqrt{6}}
\left[ 1-\frac{2n+1}{(2n \mu)^{2n/(2n+1)}}
\left( 1-e^{2\phi/\sqrt{6}} \right) \right]^{-2n/(2n+1)}
\left( 1-e^{2\phi/\sqrt{6}} \right)^{-1/(2n+1)}\,.
\end{eqnarray}
In the deep matter-dominated epoch in which
the condition $R/R_c \gg 1$ is satisfied,
the field $\phi$ is very close to zero.
For $n$ and $\mu$ of the order of unity,
$|\lambda|$ is much larger than unity during this stage.
Hence the matter era is realized by the instantaneous
fixed point (d).
As $R/R_c$ gets smaller, $|\lambda|$ decreases to
the order of unity.
If the solutions reach the point
$\lambda=4Q=-4/\sqrt{6}$ and satisfy the stability condition
${\rm d} \lambda/{\rm d}F<0$ the final attractor
corresponds to the de-Sitter fixed point (e).

For the theories with general couplings $Q$,
let us consider the following scalar-field potential
\begin{eqnarray}
\label{model}
V(\phi)=V_0 \left[ 1-C (1-e^{-2Q\phi})^p \right]
\qquad (V_0>0,~C>0,~0<p<1)\,,
\end{eqnarray}
as a natural generalization of Eq.~(\ref{fRpo}).
The slope of the potential is given by
\begin{eqnarray}
\label{lambda}
\lambda=
\frac{2C\,p\,Q e^{-2Q \phi} (1-e^{-2Q\phi})^{p-1}}
{1-C(1-e^{-2Q\phi})^p}\,.
\end{eqnarray}
When $Q>0$, the potential energy decreases from $V_0$
as $\phi$ increases from 0.
On the other hand, if $Q<0$, the potential energy
decreases from $V_0$ as $\phi$ decreases from 0.
In both cases we have $V(\phi) \to V_0(1-C)$
in the limits $\phi \to \infty$ (for $Q>0$) and
$\phi \to -\infty$ (for $Q<0$).

In the model (\ref{model}) the field is stuck around the value
$\phi=0$ during the deep radiation and matter epochs.
In these epochs one has $R \simeq \rho_{m}/F$ from
Eqs.~(\ref{be1}), (\ref{be2}) and (\ref{Ricci}) by noting
that $V_0$ is negligibly small compared to
$\rho_{m}$ or $\rho_{\rm rad}$.
Using Eq.~(\ref{be5}), we obtain the relation
$V_{,\phi}+Q\rho_{m} \simeq 0$.
Hence, in the high-curvature region the field $\phi$
evolves along the instantaneous minimum given by
\begin{eqnarray}
\label{phim}
\phi_m \simeq \frac{1}{2Q} \left(
\frac{2V_0pC}{\rho_m} \right)^{\frac{1}{1-p}}\,.
\end{eqnarray}

We stress here that a range of minima appears
depending upon the large energy density $\rho_m$
of the non-relativistic matter.
As long as the condition $\rho_m \gg V_0 pC$ is
satisfied, we have $|\phi_m| \ll 1$ from
Eq.~(\ref{phim}).

Since from Eq.~(\ref{lambda}) $|\lambda| \gg 1$ for field values around $\phi=0$,
the instantaneous fixed point (d) can represent the matter-dominated epoch
provided that $|Q| \ll |\lambda|$.
The deviation from Einstein gravity manifests itself when
the field begins to evolve towards the end of the matter era.
The variable $F=e^{-2Q\phi}$ decreases in time irrespective
of the sign of the coupling strength and hence $0<F<1$.
This decrease of $F$ is crucial to the
divergent behavior of the equation of state of DE,
as we will see in Sec.~\ref{secSN}.

The de-Sitter solution corresponds to
$\lambda=4Q$, i.e.,
\begin{eqnarray}
\label{Cvalue}
C=\frac{2}{(1-F_1)^{p-1}
\left[ 2+(p-2)F_1 \right]}\,,
\end{eqnarray}
where $F_1$ is the value of $F$ at the point (e).
Provided that the solution of this equation
exists in the region $0<F_1<1$, for given
values of $C$ and $p$,
the de-Sitter point exists.
{} From Eq.~(\ref{lambda}) we obtain
\begin{eqnarray}
\frac{{\rm d}\lambda}{{\rm d}\phi}=
-\frac{4CpQ^2F(1-F)^{p-2}[1-pF-C(1-F)^p]}
{[1-C(1-F)^p]^2}\,.
\end{eqnarray}

When $0<C<1$, one can easily show that
the function $g(F) \equiv 1-pF-C(1-F)^p$
is positive in the region $0<F<1$
giving ${\rm d}\lambda/{\rm d}\phi<0$.
Hence, the conditions for a stable de-Sitter
point is automatically satisfied.
In this case the solutions approach the de-Sitter attractor after the
end of the matter era.

When $C>1$, the function $g(F)$ becomes
negative for values of $F$ that are smaller than the critical value
$F_c~(<1)$.
The de-Sitter point (e) is stable under
the condition $1-pF_1>C(1-F_1)^p$.
Using Eq.~(\ref{Cvalue}) we find that this
stability condition translates to
\begin{eqnarray}
\label{F1con}
F_1>\frac{1}{2-p}\,.
\end{eqnarray}
If this condition is violated, the solutions choose
another stable fixed point as an attractor.
In $f(R)$ gravity, for example, the solutions can reach
the stable accelerated point (d) characterized b,
$m=-r-1$ and $(\sqrt{3}-1)/2<m<1$ \cite{AGPT},
where $m \equiv Rf_{,RR}/f_{,R}$ and
$r \equiv -Rf_{,R}/f$.

\begin{figure}
\includegraphics[height=3.4in,width=3.4in]{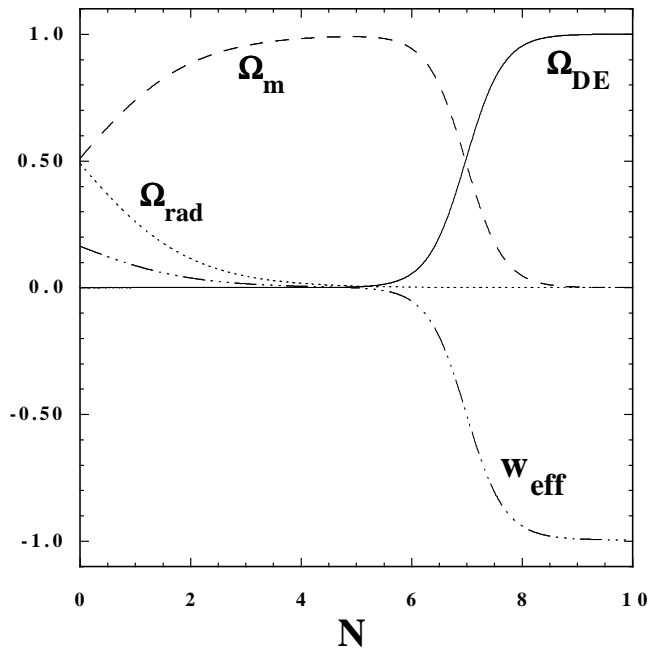}
\caption{
The evolution of $\Omega_{\rm DE}$, $\Omega_m$,
$\Omega_{\rm rad}$ and $w_{\rm eff}$
for the model (\ref{model}) with
parameters $Q=0.01$, $p=0.2$
and $C=0.7$ and initial conditions
$x_1=0$, $x_2=2.27 \times 10^{-7}$,
$x_3=0.7$ and $x_4-1=-5.0 \times 10^{-13}$.}
\label{evo}
\end{figure}

In summary, when $0<C<1$, the matter point (d) can be
followed by the stable de-Sitter solution (e) for
the model (\ref{model}).
In Fig.~\ref{evo} we plot the evolution of
$\Omega_{\rm DE}$, $\Omega_m$,
$\Omega_{\rm rad}$ and $w_{\rm eff}$
for $Q=0.01$, $p=0.2$ and $C=0.7$.
Beginning from the epoch of matter-radiation equality,
the solutions first dwell around the matter point (d)
with $w_{\rm eff} \simeq 0$
and finally approach the de-Sitter attractor (e)
with $w_{\rm eff} \simeq -1$.
We have also numerically confirmed that
$\lambda$ is initially much larger than unity and
eventually approaches the value $\lambda=4Q$.

\section{Local gravity constraints}
\label{secLGC}

In this section we shall study the local gravity constraints (LGC)
for the scalar-tensor theories given by the action (\ref{action2}).
In the absence of the potential $V(\phi)$
the Brans-Dicke parameter $\omega_{\rm BD}$
is constrained to be $\omega_{\rm BD}>4.0 \times 10^4$
from solar-system experiments \cite{hoyle}.
Note that this bound also applies to the case
of a nearly massless field with the potential $V(\phi)$
in which the Yukawa correction $e^{-Mr}$ is close to unity
(where $M$ is the scalar field mass and $r$ is
an interaction length).
Using the bound $\omega_{\rm BD}>4.0 \times 10^4$
in Eq.~(\ref{BD}), we find
\begin{eqnarray}
|Q|<2.5 \times 10^{-3}\qquad
({\rm for~the~massless~case}).
\end{eqnarray}
This is a strong constraint under which the
cosmological evolution for such theories is difficult
to be distinguished from the $Q=0$ case.

Let us then consider the case in which the mass $M$
of the field $\phi$ is sufficiently heavy so that the
interaction range of the field ($\sim 1/M$) becomes
short so as to satisfy LGC.
In the context of $f(R)$ gravity ($Q=-1/\sqrt{6}$)
it was in fact shown that the LGC can be satisfied
by constructing models in which $M$ is large enough
in a high-dense region where local gravity experiments
are carried out \cite{Hu07,Star07,Appleby,Tsuji08,TUT}.
In these models the mass $M$ tends to become lighter
with the decrease of the Ricci scalar $R$ towards
the present epoch.
In what follows, we shall construct viable
models, based on scalar-tensor theories whose couplings $Q$
are of order unity, which are consistent with LGC.

\subsection{Chameleon mechanisms}

In Refs.~\cite{Nava,Max,Hu07,CT} it was explicitly shown
that in $f(R)$ gravity a spherically symmetric body
forms a thin-shell inside the body through a
chameleon mechanism \cite{KW1,KW2}.
Generally this happens in a non-linear regime where the
mass $M$ of a scalar-field degree of freedom is heavy
so that the usual linear analysis based on the inequality,
$|\delta R| \ll |R^{(0)}|$, is invalid (where $\delta R$ is a
perturbation about a background value $R^{(0)}$).
In what follows we shall briefly
review the chameleon mechanism for the theory
given in Eq.~(\ref{action2}) and then
place constraints on viable models
consistent with LGC.

Let us consider the Einstein frame action (\ref{SE}).
The variation of this action with respect to $\phi$
leads to the following equation of motion
\begin{eqnarray}
\label{phieq}
\tilde{\square} \phi-U_{,\phi}
=-Q\tilde{T}\,,
\end{eqnarray}
where, $\tilde{T}=e^{4Q \phi}T$ and
$T=g^{\mu \nu}T_{\mu \nu}$, with $T_{\mu \nu}$
being energy momentum tensor of the matter
in the Jordan frame.
We take a spherically symmetric spacetime
with a radius $\tr$ from the center of symmetry.
In this setup Eq.~(\ref{phieq}) becomes

\begin{eqnarray}
\label{dreq}
\frac{{\rm d}^2 \phi}{{\rm d} \tr^2}+
\frac{2}{\tr} \frac{{\rm d}\phi}{{\rm d}\tr}=
\frac{{\rm d}U_{\rm eff}}{{\rm d}\phi}\,,
\end{eqnarray}
where
\begin{eqnarray}
\label{Ueff}
U_{\rm eff}(\phi)=U(\phi)+
e^{Q \phi}\rho^*\,.
\end{eqnarray}
Here, $\rho^*$ is related with the energy density $\rho \equiv -T$
in the Jordan frame via the relation $\rho^*=e^{3Q \phi}\rho$,
which is conserved in the Einstein frame \cite{KW2}
({i.e.}, $\rho^* \tr^3=$constant).

We consider a configuration in which the spherically symmetric body has a constant
density $\rho^*=\rho_A^*$ inside the body ($\tr<\tr_c$)
and that the energy density outside the body ($\tr>\tr_c$)
is given by $\rho^*=\rho_B^*~(\ll \rho_A^*)$. Then the mass of this body is
given by $M_c=(4\pi/3)\tr_c^3\rho_A^*=(4\pi/3)r_c^3\rho_A$.
Let us denote the field value at the minimum of the effective
potential $U_{\rm eff}(\phi)$ corresponding to the density $\rho_A^*$
($\rho_B^*$) by $\phi_A$ ($\phi_B$).  That is, they are given by
\begin{eqnarray}
\label{Ucon}
U_{,\phi} (\phi_A)+Q e^{Q \phi_A}\rho_A^*=0\,,~~{\rm and}~~
U_{,\phi} (\phi_B)+Q e^{Q \phi_B}\rho_B^*=0\,,
\end{eqnarray}
respectively.
Under the condition $\rho_A^* \gg \rho_B^*$
the mass squared $m_A^2 \equiv U_{{\rm eff}}''(\phi_A)$
is much larger than
$m_B^2 \equiv U_{{\rm eff}}''(\phi_B)$.

In solving for a static spherically symmetric
field configuration, we impose
the boundary conditions
${\rm d}\phi (\tr=0)/{\rm d}\tr=0$ and
$\phi (\tr\to \infty)=\phi_B$, so that
the field $\phi$ is non-singular at $\tr=0$ and
that the force on a test body vanishes at a sufficiently
large distance.
Then $\phi$ starts to roll down the potential where the term
${\rm d}U_{\rm eff}/{\rm d}\phi$ becomes important.

If the field value at the center $\phi(\tr=0)$ is close enough
to the equilibrium value $\phi_A$ with
$|\phi(\tr=0)-\phi_A| \ll |\phi_A|$, the
thin-shell solution is realized \cite{KW2}.
In this case the field does not move away from $\phi(\tr=0)$
practically up to a radius $\tr_1$ which satisfies
$\Delta \tr_c/\tr_c \equiv (\tr_c-\tr_1)/\tr_c \ll 1$.
At $\tr=\tr_1$,
the field starts to roll down the potential and
 we find, $|U_{,\phi} (\phi)| \ll
|Q e^{Q \phi}\rho_A^*|$ for $\tr_1<\tr<\tr_c$.
Under the condition $|Q \phi_A| \ll 1$ (as we will confirm later)
the r.h.s. of Eq.~(\ref{dreq}) is approximately
given by ${\rm d}U_{\rm eff}/{\rm d}\phi \simeq Q \rho_A^*$.
Outside the body ($\tr>\tr_c$) the gradient
energies on the l.h.s. of Eq.~(\ref{dreq}) become important
because the energy density drops down from $\rho_A^*$
to $\rho_B^*$.
Taking into account the mass term $m_B$ of the effective
potential $U_{\rm eff}$, one has
${\rm d}U_{\rm eff}/{\rm d}\phi=m_B^2 (\phi-\phi_B)$
on the r.h.s. of Eq.~(\ref{dreq}).

We match the solutions of Eq.~(\ref{dreq}) at $\tr=\tr_c$
with the boundary conditions $\phi=\phi_A$,
${\rm d}\phi/{\rm d}\tr=0$ at $\tr=\tr_1$ and
$\phi (\tr \to \infty)=\phi_B$.
Then the following solution is obtained
in the region $\tr>\tr_c$ \cite{KW2,CT}.
\begin{eqnarray}
\label{sol2}
\phi(\tr) \simeq
-\frac{Q M_c}{4\pi} \left[1-
\left(\frac{\tr_1}{\tr_c} \right)^3 \right]
\frac{e^{-m_B(\tr-\tr_c)}}{\tr}+\phi_B\,,
\end{eqnarray}
where
\begin{eqnarray}
\label{rat}
\left( \frac{\tr_1}{\tr_c} \right)^2
&\simeq&1-\frac{\phi_B-\phi_A}{3Q \Phi_c}\,,
\qquad
\Phi_c \equiv \frac{M_c}{8\pi \tr_c}
=\frac{GM_c}{\tr_c}\,.
\end{eqnarray}
In deriving the relation (\ref{sol2}),
we assumed the condition $m_B\tr_c \ll 1$.
Since we are in the thin-shell regime,
we obtain the following relation from Eq.~(\ref{rat}).
\begin{eqnarray}
\label{delrc}
\frac{\Delta \tr_c}{\tr_c}
\simeq \frac{\phi_B-\phi_A}{6Q \Phi_c}\,.
\end{eqnarray}
Then the solution outside
the body ($\tr>\tr_c$) is given by
\begin{eqnarray}
\label{phir1}
\phi(\tr) \simeq -\frac{Q_{\rm eff}}{4\pi}
\frac{M_c e^{-m_B(\tr-\tr_c)}}{\tr}+\phi_B\,,
\quad
{\rm where}
\quad
Q_{\rm eff} \equiv 3Q
\frac{\Delta \tr_c}{\tr_c}\,.
\end{eqnarray}
Thus, when the body has a thin-shell,
the effective coupling $|Q_{\rm eff}|$
in the thin-shell becomes much smaller than unity,
even if $|Q|$ itself is of the order of 1.

If the field value at $\tr=0$ is not close to $\phi_A$
({i.e.}, $|\phi(\tr=0)-\phi_A| \gtrsim |\phi_A|$),
the field rapidly rolls down the potential at $\tr_1 \simeq 0$.
Setting $\tr_1=0$ in Eq.~(\ref{sol2}), we obtain the solution
(\ref{phir1}) with $Q_{\rm eff}$ replaced by $Q$.
This is the thick-shell regime in which the effective coupling
is not small as to satisfy the LGC.

The presence of the fifth force interaction mediated by
the field $\phi$ leads to a modification to
the spherically symmetric metric.
Under the weak field approximation, the spherically
symmetric metric in the Jordan frame is
given by \cite{Max,CT}
\begin{eqnarray}
{\rm d}s^2=-\left(1-\frac{2G_{\rm eff}M_c}{r}
\right) {\rm d}t^2+\left( 1+\frac{2\gamma G_{\rm eff}M_c}{r}
\right){\rm d}r^2+r^2({\rm d}\theta^2+\sin^2 \theta {\rm d}\phi^2)\,,
\end{eqnarray}
where the effective gravitational ``constant'' $G_{\rm eff}$
and the post-Newtonian parameter $\gamma$ are given by
\begin{eqnarray}
G_{\rm eff} \simeq G \left[
1-\frac{\sqrt{6}}{3}Q_{\rm eff}
e^{-m_B (r-r_c)} \right]\,,\qquad
\gamma \simeq
\frac{1+(\sqrt{6}Q_{\rm eff}/3)(1+m_{B} r)
e^{-m_B (r-r_c)}}{1-(\sqrt{6}Q_{\rm eff}/3)
e^{-m_B (r-r_c)}}\,.
\end{eqnarray}
Note that we have used the approximation $\tr \simeq r$
that is valid  in the region $|Q\phi| \ll 1$.

Provided that the condition $m_Br \ll 1$
holds in an environment in which local gravity
experiments are carried out, we have
$\gamma \simeq (1+\sqrt{6}Q_{\rm eff}/3)
/(1-\sqrt{6}Q_{\rm eff}/3)$.
Hence, if $|Q_{\rm eff}|$ is much smaller than unity
through the chameleon mechanism, it is possible to
satisfy the following severest solar system constraint
that comes from a time-delay effect of
the Cassini tracking \cite{Will}:
\begin{eqnarray}
\label{gamcon}
\left| \gamma-1 \right|<2.3 \times 10^{-5}\,.
\end{eqnarray}
Using the thin-shell parameter,
this bound translates into
\begin{eqnarray}
\label{delrcon}
\frac{\Delta r_c}{r_c}<
\frac{4.7 \times 10^{-6}}{|Q|}\,.
\end{eqnarray}
If the body does not have a thin-shell for $|Q|$
of the order of unity, the condition (\ref{gamcon})
is not satisfied.
In $f(R)$ gravity, for example,  we have $\gamma=1/2$
for $Q_{\rm eff}=Q=-1/\sqrt{6}$.

\subsection{Solar system constraints}

In $f(R)$ gravity, the models (\ref{fRmodel}) can satisfy LGC because the mass
$M$ of the field potential (\ref{fRpo}) is sufficiently heavy in the high-density
region whose Ricci scalar $R$ is much larger than $R_c$.
Since the field mass $M_{\phi}$ inside the body is much heavier than that outside
the body, most of the volume element within the core does not
contribute to the field profile at $r>r_c$ except for the thin-shell regime
around the surface of the body
(note that this contribution is proportional to $e^{-M\ell}$,
where $\ell$ is a distance from the volume element to
a point outside the body).
In the case of general couplings $Q$, the models
presented in Eq.~(\ref{model}) can be compatible with LGC.
Under the condition $|Q \phi| \ll 1$, one has
$U_{,\phi} \simeq -2V_0 QpC(2Q\phi)^{p-1}$
for the potential $U=V/F^2$ in the
Einstein frame.
Then from Eq.~(\ref{Ucon}) we obtain the field values
at the potential minima inside/outside the body:
\begin{eqnarray}
\phi_A \simeq \frac{1}{2Q} \left(
\frac{2V_0\,p\,C}{\rho_A} \right)^{\frac{1}{1-p}}\,,
\quad
\phi_B \simeq \frac{1}{2Q} \left(
\frac{2V_0\,p\,C}{\rho_B} \right)^{\frac{1}{1-p}}\,,
\label{phiB}
\end{eqnarray}
which satisfy $|\phi_A| \ll |\phi_B|$.
Note that these are analogous to the field value $\phi_m$
derived in Eq.~(\ref{phim}) in the cosmological setting.
In order to realize the accelerated expansion at the
present
epoch, $V_0$ needs to be roughly the same order as the square
of the present Hubble parameter $H_0$, so we have $V_0 \sim H_0^2
\sim \rho_0$, where $\rho_0 \simeq 10^{-29}$ g/cm$^3$
is the present cosmological density.
Note that the baryonic/dark matter density in our galaxy
corresponds to $\rho_B \simeq 10^{-24}$ g/cm$^3$.
This then shows that the conditions, $|Q\phi_A| \ll 1$
and $|Q\phi_B| \ll 1$ are in fact satisfied provided that
$C$ is not much larger than unity.

The field mass squared $M_\phi^2 \equiv {\rm d}^2 U/{\rm d} \phi^2$
at $\phi=\phi_A$ is approximately given by
\begin{eqnarray}
\label{Mphi}
M_\phi^2(\phi_A) \simeq
\frac{1-p}{(2^p\,pC)^{1/(1-p)}}Q^2
\left( \frac{\rho_A}{V_0}
\right)^{\frac{2-p}{1-p}} V_0\,.
\end{eqnarray}
This means that $M_\phi(\phi_A)$ can be much larger than
$H_0$ due to  the condition $\rho_A \gg V_0$.
Therefore, while the mass $M_{\phi}$ is not different from
the order of $H_0$ on cosmological scales, it increases in the
regions with a higher energy density.

Let us place constraints on model parameters by using
the solar system bound (\ref{delrcon}).
In so doing, we shall consider the case where
the solutions finally approach the de-Sitter point (e).
Since we have $\Delta r_c/r_c \simeq \phi_B/(6Q\Phi_c)$
with $\phi_B$ given in Eq.~(\ref{phiB}), the bound
(\ref{delrcon}) translates into
\begin{eqnarray}
\label{delrcon2}
\left( 2V_0pC/\rho_B \right)^{1/(1-p)}
<1.2 \times 10^{-10} |Q|\,,
\end{eqnarray}
where we have used the value
$\Phi_c=2.12 \times 10^{-6}$ for the Sun.
At the de-Sitter point (e), one has
$3F_1H_1^2=V_0[1-C(1-F_1)^p]$ with
$C$ given in Eq.~(\ref{Cvalue}).
Hence, we obtain the following relation
\begin{eqnarray}
V_0=3H_1^2 \frac{2+(p-2)F_1}{p}\,.
\end{eqnarray}
Substituting this in Eq.~(\ref{delrcon2}) we find
\begin{eqnarray}
\left( \frac{R_1}{\rho_B} \right)^{1/(1-p)}
(1-F_1) <1.2 \times 10^{-10}\,|Q|\,,
\end{eqnarray}
where $R_1=12H_1^2$ is the Ricci scalar
at the de-Sitter point.
Since the term $(1-F_1)$ is smaller than one half
from the condition (\ref{F1con}) we obtain
the inequality $(R_1/\rho_B)^{1/(1-p)}
<2.4 \times 10^{-10} |Q|$.
We assume that
$R_1$ is of the order of
the present cosmological density
$\rho_0=10^{-29}$ g/cm$^3$.
Taking the baryonic/dark matter density to be
$\rho_B=10^{-24}$ g/cm$^3$ outside the Sun
we obtain the following bound
\begin{eqnarray}
\label{solar}
p>1-\frac{5}{9.6-{\rm log}_{10}\,|Q|}\,.
\end{eqnarray}
For $|Q|=10^{-2}$ and $|Q|=10^{-1}$
this gives $p>0.57$ and $p>0.53$ respectively.
The above bound corresponds to
$p>0.50$ for the case of $f(R)$ gravity, which translates
into the condition $n>0.5$ in Eq.~(\ref{fRpo}).
This agrees with the result found in Ref.~\cite{CT}.

\subsection{Equivalence principle constraints}

Let us proceed to the constraints from a possible
violation of the equivalence principle (EP).
Under the condition that the neighborhood of
the Earth has a thin-shell, the tightest bound
comes from solar system tests of the EP that makes
use of the free-fall accelerations of the Moon
($a_{\rm Moon}$) and the Earth ($a_{\oplus}$)
toward the Sun \cite{KW2,CT}.
The bound on the differences between the two
accelerations is \cite{Will}
\begin{eqnarray}
\label{EPsolar}
2\frac{|a_{\rm Moon}-a_{\oplus}|}
{a_{\rm Moon}+a_{\oplus}}<10^{-13}\,.
\end{eqnarray}

Since the acceleration induced
by a fifth force with the field profile $\phi(r)$
and the effective coupling is given by
$a_{\rm fifth}=|Q_{\rm eff}\phi(r)|$
we obtain \cite{KW2}
\begin{eqnarray}
a_{\oplus} = \frac{GM_{\odot}}{r^2}
\left[ 1+3 \left(\frac{\Delta r_{\oplus}}
{r_{\oplus}}\right)^2 \frac{\Phi_{\oplus}}{\Phi_{\odot}}
\right]\,,\quad
a_{\rm Moon} =
\frac{GM_{\odot}}{r^2}
\left[ 1+3 \left(\frac{\Delta r_{\oplus}}
{r_{\oplus}}\right)^2 \frac{\Phi_{\oplus}^2}
{\Phi_{\odot}\Phi_{\rm Moon}}\right]\,,
\end{eqnarray}
where $\Phi_{\odot} \simeq 2.1 \times 10^{-6}$,
$\Phi_{\oplus} \simeq 7.0 \times 10^{-10}$ and
$\Phi_{\rm Moon} \simeq 3.1 \times 10^{-11}$,
are the gravitational potentials of the Sun,
the Earth and the Moon, respectively.
Note that $\Delta r_{\oplus}/r_{\oplus}$ is the
thin-shell parameter of the Earth.
From the bound (\ref{EPsolar}), this is constrained to be
\begin{eqnarray}
\frac{\Delta r_{\oplus}}{r_{\oplus}}
<\frac{8.8 \times 10^{-7}}{|Q|}\,.
\end{eqnarray}
Note also that the thin-shell condition for the neighborhood
outside the Earth provides the same order of the upper
bound for, $\Delta r_{\oplus}/r_{\oplus}$ \cite{CT}.

Taking a similar procedure as in the case of
the solar system constraints discussed
above (using the value $R_1=10^{-29}$ g/cm$^3$
and $\rho_B=10^{-24}$ g/cm$^3$),
we obtain the following bound:
\begin{eqnarray}
\label{EP}
p>1-\frac{5}{13.8-{\rm log}_{10}\,|Q|}\,.
\end{eqnarray}
This is tighter than the bound (\ref{solar}).
When $|Q|=10^{-2}$ and $|Q|=10^{-1}$
we have $p>0.68$ and $p>0.66$, respectively.
In the case of $f(R)$ gravity
the above bound corresponds to $p>0.65$
which translates to $n>0.9$
for the potential (\ref{fRpo}).

In summary, the LGC can be satisfied under the condition
(\ref{EP}) for the potential (\ref{model}).

\subsection{General properties for models consistent with LGC}

In this subsection we shall consider the
general properties of scalar-tensor theories
consistent with LGC, without specifying the form of the field potential.
In order to satisfy the LGC we require that $|\phi_B-\phi_A|$
is much smaller than $|Q\Phi_c|$ from Eq.~(\ref{delrc}).
Since there is a gap between the energy densities inside and outside
of the spherically symmetric body we have,
$|\phi_B-\phi_A| \simeq |\phi_B|$,
which implies $|\phi_B| \ll |Q\Phi_c|$.
The gravitational potential $\Phi_c$ is very much smaller than
unity in settings where local gravity experiments
are carried out, hence this yields the constraint $|\phi_B| \ll 1$.
Cosmologically this means that $|\phi|$ is much smaller than
unity during matter/radiation epochs.
When $|Q| \gg 1$ the condition
$|\phi_B| \ll 1$ is not necessarily ensured, but those cases
are excluded by the constraints from density perturbations
unless the model is very close to the $\Lambda$CDM model
(as we shall see later).
In the following we shall consider
the theories with $|Q| \lesssim 1$.

In the region $|\phi| \ll 1$ (i.e., $F \simeq 1$),
the derivative terms are negligible in Eq.~(\ref{be5})
and the field stays at instantaneous minima given by
\begin{eqnarray}
\label{Vphicon}
V_{,\phi}+QFR=0\,,
\end{eqnarray}
in the late radiation-dominated and matter-dominated eras.
The condition (\ref{Vphicon}) translates into
$\lambda/Q=\rho_m/V$ which means that $\lambda/Q \gg 1$
in the radiation and matter epochs.
This is in fact consistent with the condition
$|w_{\rm eff}| =|2Q/\lambda| \ll 1$
for the existence of a viable matter point (d).
If the de-Sitter point (e) is stable, the solutions finally
approach the minimum given by (\ref{Vphicon}), i.e., $\lambda/Q=4$.

\begin{figure}
\includegraphics[height=3.7in,width=4.0in]{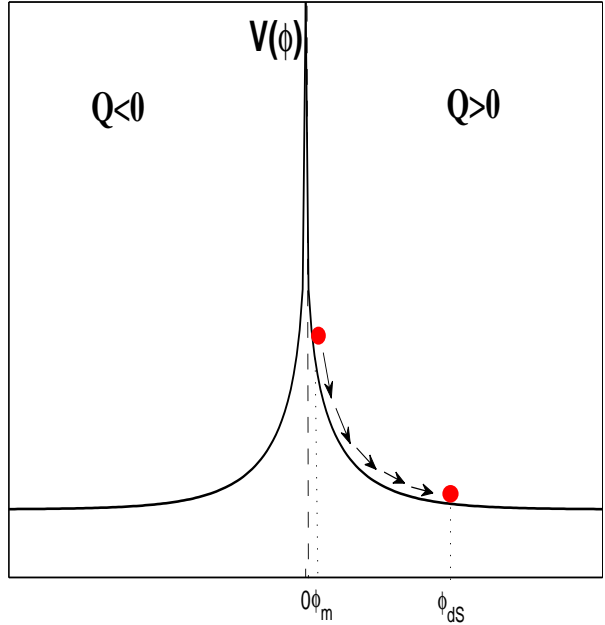}
\caption{
This illustration describes a field potential $V(\phi)$ that
is consistent with LGC. For a coupling $Q$ that is positive
(negative) the potential evolves in the region
$\phi \ge 0$ ($\phi \le 0$).
In the figure $\phi_m$ represents the field value during the
radiation/matter eras, which instantaneously changes in time.
The field value $\phi_{\rm dS}$ corresponds to the one
at the de-Sitter point.
Note that both $\phi_m$ and $\phi_{\rm dS}$ are sustained by the
presence of the coupling $Q$ having potential minima
characterized by the condition (\ref{Vphicon}).
In the early stages of the cosmological evolution, the mass $M$
of the field $\phi$ is heavy for consistency with LGC.
This mass gradually gets smaller as the system approaches
the de-Sitter point.}
\label{poten}
\end{figure}

The sign of $\lambda$ needs to be the same as
that of $Q$ in order to realize the above cosmological trajectory.
When $Q>0$, we require $\lambda=-V_{,\phi}/V>0$, i.e.,
$V_{,\phi}<0$, which means that the
field $\phi$ evolves along the potential
toward larger positive values from $\phi \simeq 0$.
When $Q<0$ the field evolves towards smaller
negative values from $\phi \simeq 0$.

We illustrate such potentials
in Fig.~\ref{poten}.
Since the ratio $\lambda/Q$ decreases from the radiation/matter epochs
to the de-Sitter epoch, the derivative ${\rm d}\lambda/{\rm d}\phi$
is negative irrespective of the sign of $Q$.
We recall that in this case the stability of the de-Sitter point (e)
is also ensured.
Since ${\rm d}\lambda/{\rm d}\phi=\lambda^2 -V_{,\phi \phi}/V$,
the mass squared
\begin{eqnarray}
\label{mass}
M^2 \equiv V_{,\phi \phi}\,,
\end{eqnarray}
is required to be positive to
satisfy the condition ${\rm d}\lambda/{\rm d}\phi<0$.
Moreover, the mass $M$ needs to be heavy enough
in order to satisfy the condition
$M^2 >\lambda^2 V$ in radiation/matter epochs.
The model (\ref{model}) provides a representative example
which satisfies all the requirements discussed above.

It is worth mentioning that for the models that satisfy
LGC, the quantity $F=e^{-2Q\phi}$ in the matter/radiation eras
is larger than its value at the de-Sitter point.
It is this property which
leads to an interesting observational signature
for the DE equation of state, as we shall see in the next section.

\section{The equation of state of dark energy}
\label{secSN}

In scalar-tensor DE models, a meaningful definition
of energy density and pressure of DE requires some care.
In this section, following Ref.~\cite{DEeos},
we shall discuss the evolution of
the equation of state of DE, which could provide
comparisons with observations.
In the absence of radiation, Eqs.~(\ref{be1}) and (\ref{be2})
can be written as
\begin{eqnarray}
\label{mofrw}
3F_0H^2 &=& \rho_{\rm DE}+\rho_{m}\,, \\
\label{mofrw2}
-2F_0 \dot{H} &=& \rho_{\rm DE}+p_{\rm DE}+
\rho_{m}\,,
\end{eqnarray}
where the subscript ``0'' represents present values and
\begin{eqnarray}
\label{rhode}
\rho_{\rm DE} &\equiv&
\frac12 (1-6Q^2)F \dot{\phi}^2+V-3H\dot{F}
-3(F-F_0)H^2\,, \\
\label{pde}
p_{\rm DE} &\equiv& \frac12 (1-6Q^2)F \dot{\phi}^2
-V+\ddot{F}+2H\dot{F}+(F-F_0)(3H^2+2\dot{H})\,,
\end{eqnarray}
which satisfy the usual conservation equation
\begin{eqnarray}
\dot{\rho}_{\rm DE}+3H(\rho_{\rm DE}
+p_{\rm DE})=0\,.
\end{eqnarray}

We define the equation of state of DE to be
\begin{eqnarray}
\label{wdedef}
w_{\rm DE} \equiv \frac{p_{\rm DE}}{\rho_{\rm DE}}
=\frac{w_{\rm eff}}{1-(F/F_0)\Omega_{m}}\,,
\end{eqnarray}
where $\Omega_{m}$ and $w_{\rm eff}$
are defined in Eqs.~(\ref{con}) and (\ref{weff}), respectively.
Integrating Eq.~(\ref{be3}), we obtain
\begin{eqnarray}
\rho_m=3F_0 \Omega_{m}^{(0)}
H_0^2 (1+z)^3\,,
\end{eqnarray}
where $\Omega_{m}^{(0)}$ is the present energy
fraction of the non-relativistic matter
and $z \equiv a_0/a-1$ is the redshift.
On using Eqs.~(\ref{mofrw}) and (\ref{mofrw2}), we find
\begin{eqnarray}
w_{\rm DE}=-\frac{3r-(1+z)({\rm d}r/{\rm d} z)}
{3r-3\Omega_{m}^{(0)}(1+z)^3}\,,
\end{eqnarray}
where $r=H^2(z)/H_0^2$.
Note that this is the same equation as the one
used in Einstein gravity \cite{CST}.
By defining the energy density $\rho_{\rm DE}$
and the pressure $p_{\rm DE}$ as given in Eqs.~(\ref{rhode})
and (\ref{pde}), the resulting DE equation of state
$w_{\rm DE}$ agrees with the usual expression
which can be used to confront the models
with SNIa observations.

{}From Eq.~(\ref{wdedef}) we find that $w_{\rm DE}$
becomes singular at the point $\Omega_{m}=F_0/F$.
This happens for models in which $F$
increases from its present value $F_0$ as
we go back in time.
{}From Eq.~(\ref{conf_factor}) it is clear that $F$ decreases
in time for $Q\dot{\phi}>0$.
We note that even when the system crosses the point
$\Omega_{m}=F_0/F$ physical quantities such as the
Hubble parameter remain to be continuous.

The models (\ref{model}) satisfy this condition regardless
of the sign of $Q$, which means that the divergent behavior
of $w_{\rm DE}$ indeed occurs.
We recall that in the context of
$f(R)$ gravity ($Q=-1/\sqrt{6}$) the models
$f(R)=R-\mu^{2(n+1)}/R^n$ ($n>0$)
correspond to a scalar field potential that decreases toward
larger $\phi$, i.e., $\dot{\phi}>0$ \cite{Capo}.
Hence, the divergence of $w_{\rm DE}$ does not occur
in such models because of the decrease of
$F$ toward the past.

\begin{figure}
\includegraphics[height=3.6in,width=3.6in]{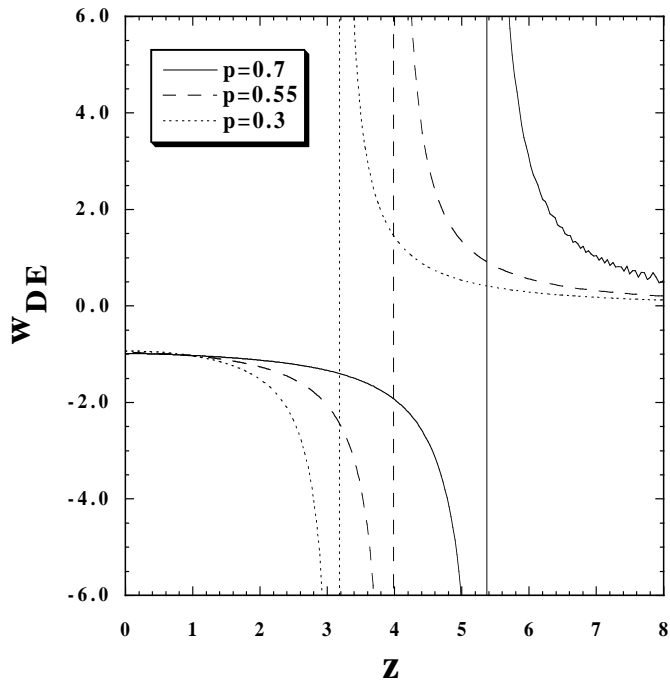}
\caption{
Figure depicting the evolution of $w_{\rm DE}$ for $Q=0.1$ and $C=0.95$
with three different values of $p$ ($0.3, 0.55, 0.7$).
The redshift $z_c$ at which the divergence of $w_{\rm DE}$
occurs decreases for smaller $p$. }
\label{wde}
\end{figure}

For the models that satisfy $|\lambda| \gg 1$
initially such that $|\lambda|$ decreases with time,
the solutions are in the regime around the instantaneous
fixed point (d) during the matter era and finally approach either
the scalar-field dominated point (c) or the de-Sitter
point (e). In Fig.~\ref{wde} we plot the evolution of $w_{\rm DE}$
for the case $Q=0.1$ and $C=0.95$, with
three different values of $p$.
In these cases the final attractor corresponds to the
de-Sitter point (e) satisfying the relation $\lambda=0.4$.
During the deep matter era the solutions evolve along the
``instantaneous'' fixed point (d) with
$\Omega_{m}$ close to 1 (because $\lambda \gg 1$).
After $\lambda$ decreases to the order of unity,
the solutions approach the de-Sitter solution (e)
with $\Omega_{m}=0$ and
$w_{\rm DE}=w_{\rm eff}=-1$.

Figure \ref{wde} clearly shows that $w_{\rm DE}$
exhibits a divergence at a redshift $z_c$ that depends
on the values of $p$.
When $p=0.3$, for example, the divergence occurs
around the redshift, $z_c=3$.
For compatibility with LGC we require $p>0.53$,
from solar system constraints, and $p>0.66$ from
EP constraints, as we showed in the previous section.
In those cases the critical redshift $z_c$
gets larger, which is out of the observational range of
current SNIa observations.
Nevertheless, the equation of state $w_{\rm DE}$
shows a peculiar evolution that changes from $w_{\rm DE}<-1$
to $w_{\rm DE}>-1$ at a redshift around $z_c={\cal O}(1)$.
This cosmological boundary crossing,
similar to the divergence of $w_{\rm DE}$, is attributed
to the fact that $F$ increases as
we go back to the past.
It is worth noting that this is
a common feature among viable models that are consistent
with LGC, as we have illustrated in the previous section.

Note that in the limit $Q \to 0$ the potential $V(\phi)$
approaches a constant value $V(\phi) \to V_0 (1-C)$.
Hence, the models are hardly distinguishable from
the $\Lambda$CDM model.
In these cases the critical redshift $z_c$ also goes to
infinity. Thus, the effect of modified gravity is more apparent
for larger $|Q|$ and smaller $p$.
In $f(R)$ gravity, for example, the model given by
Eq.~(\ref{fRpo}) can give rise to the redshift
$z_c$ as close as a few \cite{Tsuji08}
while satisfying the LGC ($p>0.65$).
These cases are particularly interesting to
place tight bounds on model parameters
from future high-precision observations.

\section{Matter density perturbations}
\label{secmatter}

In this section we discuss the evolution of matter density
perturbations and the resulting spectra for scalar-tensor theories.
Considering as our background spacetime the
flat FLRW metric, the perturbed metric including the scalar
metric perturbations $\Phi$ and $\Psi$ in the longitudinal gauge
is given by \cite{metper}
\begin{eqnarray}
\rd s^2=-(1+2\Phi) \rd t^2
+a^2(t) (1-2\Psi)
\rd x^i \rd x^j\,.
\end{eqnarray}
In the following we shall neglect radiation and consider
only the pressureless matter.
The components of the energy momentum tensor
of the pressureless matter are given by
\begin{eqnarray}
T^{0}_{0}=-(\rho_m+\delta \rho_m)\,,\quad
T^0_{i}=-\rho_mv_{m,i}\,,
\end{eqnarray}
where $v_m$ is related to the velocity potential $V$
through $v_m=-(V-b)$.
In the Fourier space matter perturbations
satisfy the following equations of motion \cite{HN,Boi,TUT}:
\begin{eqnarray}
\label{ma1d}
& & \Phi=\dot{v}\,, \\
\label{ma2d}
& &  \left( \delta \rho_m/\rho_m
\right)^{\cdot}=3\dot{\Psi}
-\frac{k^2}{a^2}v\,,
\end{eqnarray}
where $v \equiv av_m$ is a covariant
velocity perturbation and
$k$ is a comoving wavenumber.
We shall introduce the following gauge-invariant
density contrast:
\begin{equation}
\label{deltam}
\delta_m = \frac{\delta \rho_m}{\rho_m}+3Hv\,.
\end{equation}
The evolution equation for $\delta_m$
is then given by
\begin{eqnarray}
\label{mattere}
\ddot{\delta}_m+2H\dot{\delta}_m+
\frac{k^2}{a^2}\Phi=3\ddot{B}+6H\dot{B}\,,
\end{eqnarray}
where $B=Hv+\Psi$.

In Fourier space the scalar metric perturbations in scalar-tensor
theories satisfy the following equations \cite{HN}
\begin{eqnarray}
\label{per1}
& & \frac{k^2}{a^2}\Psi+3H(H\Phi+\dot{\Psi})
=-\frac{1}{2F} \Biggl[ \omega \dot{\phi} \delta
\dot{\phi}+\frac12 (\omega_{,\phi} \dot{\phi}^2
-F_{,\phi}R+2V_{,\phi})\delta \phi-3H\delta \dot{F}
+\left( 3\dot{H}+3H^2-\frac{k^2}{a^2} \right)\delta F
\nonumber \\
& &~~~~~~~~~~~~~~~~~~~~~~~~~~~~~~~~~~~~~~~
+(3H\dot{F}-\omega \dot{\phi}^2)\Phi
+3\dot{F} (H\Phi+\dot{\Psi})+\delta \rho_m
\Biggr]\,,\\
\label{per2}
& & H\Phi+\dot{\Psi}=\frac{1}{2F}
\left( \omega \dot{\phi} \delta \phi+\delta \dot{F}
-H \delta F-\dot{F}\Phi+\rho_mv \right)\,,\\
\label{per3}
& & \Psi -\Phi=\frac{\delta F}{F}\,,\\
\label{per4}
& &\delta \ddot{\phi}+\left( 3H+
\frac{\omega_{,\phi}}{\omega} \dot{\phi} \right)
\delta \dot{\phi}+\left[ \frac{k^2}{a^2}
+\left( \frac{\omega_{,\phi}}{\omega} \right)_{,\phi}
\frac{\dot{\phi}^2}{2}+
\left(\frac{2V_{,\phi}-F_{,\phi}R}{2\omega}
\right)_{,\phi} \right]\delta \phi \nonumber \\
& &=\dot{\phi}\dot{\Phi}+
\left( 2\ddot{\phi}+3H\dot{\phi}+
\frac{\omega_{,\phi}}{\omega} \dot{\phi}^2
\right) \Phi+3\dot{\phi} (H\Phi+\dot{\Psi})
+\frac{1}{2\omega} F_{,\phi} \delta R\,,
\end{eqnarray}
where $\omega=(1-6Q^2)F$ and
\begin{eqnarray}
\label{per5}
\delta R=2\left[ -3(H\Phi+\dot{\Psi})^{\cdot}
-12H(H\Phi+\dot{\Psi})+\left( \frac{k^2}{a^2}
-3\dot{H} \right)\Phi-2 \frac{k^2}{a^2}\Psi \right]\,.
\end{eqnarray}

As long as the mass $M$ defined in Eq.~(\ref{mass})
is sufficiently heavy to satisfy the conditions $M^2 \gg R$
and $M^2>\lambda^2 V$ (in order to ensure
${\rm d}\lambda/{\rm d}\phi<0$), one can approximate
$((2V_{,\phi}-F_{,\phi}R)/2\omega)_{,\phi}
\simeq M^2/\omega$ in Eq.~(\ref{per4}).
While this quantity becomes negative for $Q^2>1/6$
this does not imply that the perturbation $\delta \phi$
exhibits a negative instability.
In fact we shall illustrate below, that due to the
perturbation $\delta R$ on the r.h.s. of Eq.~(\ref{per4}),
the effective mass produced is positive.

Generally, the solution of Eq.~(\ref{per4})
consists of the sum of the matter-induced mode
$\delta \phi_{\rm ind}$ sourced by the matter
perturbation and the oscillating mode
$\delta \phi_{\rm osc}$ (scalarons \cite{Star80}), i.e.,
$\delta \phi=\delta \phi_{\rm ind}+\delta \phi_{\rm osc}$.
The oscillating mode corresponds to the solution of
Eq.~(\ref{per4}) without the matter perturbation.

In order to derive the approximate perturbation equations
on sub-horizon scales, we use the approximation
according to which the terms
containing $k^2/a^2$, $\delta \rho_{m}$, $\delta R$
and $M^2$ dominate in Eqs.~(\ref{per1})-(\ref{per4}).
This method was used in Refs.~\cite{Boi,CST,Tsuji07}
in the nearly massless case ($M^2 \lesssim H^2$).
In the context of $f(R)$ gravity this approximation
was shown in Ref.~\cite{TUT} to be extremely accurate
even in the massive case ($M^2 \gg H^2$)
as long as the oscillating
degrees of freedom do not dominate
over the matter-induced mode.

In order to extract the peculiar features of
the matter perturbations in scalar-tensor tensor theories,
let us first concentrate on the matter induced mode.
Under the above-mentioned approximation, we have
$\delta R_{\rm ind} \simeq -2 (k^2/a^2)
[\Psi+(F_{,\phi}/F)\delta \phi_{\rm ind}]$
from Eqs.~(\ref{per3}) and (\ref{per5}),
where the subscript ``ind'' represents a matter
induced mode.
Then from Eq.~(\ref{per4}) we find
\begin{eqnarray}
\label{delphi}
\delta \phi_{\rm ind} \simeq \frac{2QF}
{(k^2/a^2)(1-2Q^2)F+M^2}
\frac{k^2}{a^2}\Psi\,.
\end{eqnarray}
Using Eq.~(\ref{per1}) and (\ref{per3}) we obtain
\begin{eqnarray}
\label{phipsi}
\frac{k^2}{a^2} \Psi \simeq
-\frac{\delta \rho_m}{2F}
\frac{(k^2/a^2)(1-2Q^2)F+M^2}
{(k^2/a^2)F+M^2}\,,\qquad
\frac{k^2}{a^2} \Phi \simeq
-\frac{\delta \rho_m}{2F}
\frac{(k^2/a^2)(1+2Q^2)F+M^2}
{(k^2/a^2)F+M^2}\,.
\end{eqnarray}
In the limit $M^2/F \gg k^2/a^2$ one has
$(k^2/a^2)\Phi \simeq
-\delta \rho_m/2F \simeq -4\pi G
\delta \rho_m$, which recovers the
standard Poisson equation.
In the limit $M^2/F \ll k^2/a^2$
one has $(k^2/a^2)\Phi \simeq
-(\delta \rho_m/2F)(1+2Q^2)$,
where the effect of the coupling $Q$
becomes important.

{}From Eq.~(\ref{per2}) we find that $v$ is of
the order of $FH\Phi/\rho_m$.
Using the fact that $(k^2/a^2)\Phi$ is of the
order of $-(1/F)\delta \rho_m$ we can
estimate that $|3Hv/(\delta \rho_m
/\rho_m)| \sim (aH)^2/k^2 \ll 1$.
Hence we have $\delta_m \simeq
\delta \rho_m/\rho_m$ in
Eq.~(\ref{deltam}). Similarly the terms on
the r.h.s. of Eq.~(\ref{mattere}) can be neglected
relative to those on the l.h.s., which leads to
the following equation for matter perturbations:
\begin{eqnarray}
\label{delmap}
\ddot{\delta}_m+2H\dot{\delta}_m-4\pi G_{\rm eff}
\rho_m \delta_m \simeq 0\,,
\end{eqnarray}
where the effective ``cosmological'' gravitational
constant is given by
\begin{eqnarray}
\label{Geff}
G_{\rm eff}=\frac{1}{8\pi F}
\frac{(k^2/a^2)(1+2Q^2)F+M^2}
{(k^2/a^2)F+M^2}\,.
\end{eqnarray}
We can rewrite Eq.~(\ref{delmap}) by using
the derivative with respect to $N$:
\begin{eqnarray}
\label{delmap2}
\frac{{\rm d}^2 \delta_m}{{\rm d}N^2}+
\left( \frac12 - \frac32 w_{\rm eff} \right)
\frac{{\rm d} \delta_m}{{\rm d}N}
-\frac32 \Omega_m
\frac{(k^2/a^2)(1+2Q^2)F+M^2}
{(k^2/a^2)F+M^2}\delta_m \simeq 0\,.
\end{eqnarray}

We also define the effective gravitational potential
$\Phi_{\rm eff} \equiv (\Phi+\Psi)/2$, which is
directly linked with the Integrated-Sachs-Wolfe (ISW) effect
in CMB and the weak lensing in distant galaxies \cite{Song}.
From Eq.~(\ref{phipsi}) we obtain the relation
\begin{eqnarray}
\label{Phieff}
\Phi_{\rm eff} \simeq -\frac{a^2}{2k^2}
\frac{\rho_m}{F}\delta_m\,.
\end{eqnarray}

In order to confront models with weak lensing observations,
it is convenient to introduce the anisotropic parameter $\eta$
defined by, $\eta\equiv(\Phi-\Psi)/\Psi$ \cite{Kunz,Tsuji07}.
{}From Eq.~(\ref{phipsi}) we obtain
\begin{eqnarray}
\eta \simeq \frac{4Q^2 (k^2/a^2) F}
{(k^2/a^2)(1-2Q^2)F+M^2}\,,
\end{eqnarray}
which vanishes in the limit $M^2/F \gg k^2/a^2$
but approaches a value $\eta \to 4Q^2/(1-2Q^2)$,
in the limit $M^2/F \ll k^2/a^2$.
We also introduce another parameter $\Sigma\equiv q(1+\eta/2)$,
where $q$ is defined to be $(k^2/a^2)\Psi\equiv-(1/2)q
\rho_m \delta_m$.
We then have $\Sigma \simeq 1/F$,
which shows that the effective potential can be written as
$\Phi_{\rm eff} \simeq -(a^2/2k^2) \rho_m
\delta_m \Sigma$.
Hence, unlike the case of the Einstein gravity
the weak lensing potential in these scalar-tensor models of gravity are
affected by the changes of $\Sigma$ as well as $\delta_m$.

During the matter era the field $\phi$ sits
at the instantaneous
minima characterized by the condition (\ref{Vphicon}).
This is analogous to the situation
considered in subsection IV B where for
models (\ref{model}) the field value at the potential minimum
and the mass squared $M_\phi^2$ are given
by Eqs.~(\ref{phiB}) and (\ref{Mphi}) respectively.
Hence, we have the relations
$\phi \propto \rho_m^{\frac{1}{p-1}}$ and
$M^2 \propto M_{\phi}^2 \propto
\rho_m^{\frac{2-p}{1-p}}$
during the matter-dominated epoch.
The field $\phi$ can initially be heavy to satisfy the condition
$M^2/F \gg k^2/a^2$ for the modes relevant to the galaxy
power spectrum ($0.01h$\,Mpc$^{-1}\lesssim k \lesssim$ $0.2h$\,Mpc$^{-1}$).
Depending upon the model parameters and the mode $k$,
the mass squared $M^2$ can be smaller than $k^2/a^2$
during the matter era.

Next, let us consider the behavior of the oscillating mode.
Using Eqs.~(\ref{per1}) and (\ref{per3}) under
the condition $k^2/a^2 \gg H^2$ the gravitational
potentials for $\delta \rho_m=0$ are expressed
by $\delta \phi_{\rm osc}$.
Then from Eq.~(\ref{per5}) the perturbation $\delta R$
corresponding to the oscillating mode is given by
\begin{eqnarray}
\delta R_{\rm osc} \simeq 6Q \left( \delta
\ddot{\phi}_{\rm osc}
+3H\delta \dot{\phi}_{\rm osc}
+\frac{k^2}{a^2} \delta \phi_{\rm osc}
\right)\,.
\end{eqnarray}
Substituting this relation in Eq.~(\ref{per4}), we find
\begin{eqnarray}
\label{ddotphi}
\delta \ddot{\phi}_{\rm osc}
+3H\delta \dot{\phi}_{\rm osc}
+\left( \frac{k^2}{a^2}+\frac{M^2}{F}
\right) \delta \phi_{\rm osc} \simeq 0\,,
\end{eqnarray}
which is valid in the regimes $M^2 \gg \{R, \lambda^2 V\}$.
Equation (\ref{ddotphi}) clearly shows that
the effective mass for the oscillating mode is
positive even for $Q^2>1/6$.

In the following we shall confirm that as long as the oscillating mode
does not initially dominate over the matter-induced mode,
it remains subdominant throughout the cosmic history.
We shall discuss the two cases:
(i) $M^2/F \gg k^2/a^2$
and (ii)  $M^2/F \ll k^2/a^2$, separately.

\subsection{The case $M^2/F \gg k^2/a^2$}

In this regime the matter perturbation equation (\ref{delmap2})
reduces to the standard one in Einstein gravity.
During the matter era with, $w_{\rm eff} \simeq 0$ and
$\Omega_m \simeq 1$, we have the following solutions:
\begin{eqnarray}
\label{persol1}
\delta_m \propto a \propto t^{2/3}\,,\qquad
\Phi_{\rm eff}={\rm constant}\,.
\end{eqnarray}

For the model (\ref{model}) the matter-induced mode of
the field perturbation evolves as
$\delta \phi_{\rm ind} \propto \delta \rho_m/M^2
\propto t^{\frac{2(4-p)}{3(1-p)}}$.
When the frequency
$\omega_\phi=\sqrt{k^2/a^2+M^2/F}$ changes adiabatically
(i.e. $|\dot{\omega}_\phi/\omega_{\phi}^2| \ll 1$),
the WKB solution to Eq.~(\ref{ddotphi}) is given by
\begin{eqnarray}
\label{delphiosc}
\delta \phi_{\rm osc} \propto a^{-3/2}
\frac{1}{\sqrt{2\omega_\phi}}
\cos \left(\int \omega_\phi {\rm d}t \right)\,.
\end{eqnarray}
{}For the model (\ref{model}), in the
regime $M^2/F \gg k^2/a^2$, this oscillating mode
evolves as $\delta \phi_{\rm osc} \propto
t^{\frac{p}{2(1-p)}} \cos \left(c t^{-\frac{1}{1-p}}\right)$,
where $c$ is a constant.

Now since the background field $\phi$
during the matter era evolves as
$\phi \propto t^{\frac{2}{1-p}}$,
we find
\begin{eqnarray}
\label{delphiso}
\delta \phi/\phi
\simeq c_1 t^{2/3}
+c_2 t^{-\frac{4-p}{2(1-p)}}
\cos \left(c t^{-\frac{1}{1-p}} \right)\,.
\end{eqnarray}
This indicates that the matter-induced mode dominates
over the oscillating mode with time.
While the solution of the oscillating mode in Eq.~(\ref{delphiso})
is valid only in the WKB regime
($|\dot{\omega}_\phi/\omega_{\phi}^2| \ll 1$),
we have checked that $\delta \phi$ approaches
a constant value with oscillations
at the later stage in which the WKB
approximation is violated.
Hence, as long as the oscillating mode is not
overproduced in the early universe, it remains
sub-dominant relative to the matter-induced mode.
Note that this property also holds during
the radiation-dominated epoch.

\subsection{The case $M^2/F \ll k^2/a^2$}

In this regime the effective gravitational constant (\ref{Geff})
is given by $G_{\rm eff}=(1+2Q^2)/8\pi F$, which shows that
the effect of modified gravity becomes important.
{}From Eqs.~(\ref{delmap2}) and (\ref{Phieff}) we obtain
\begin{eqnarray}
\label{persol2}
\delta_m \propto t^{\frac{\sqrt{25+48Q^2}-1}{6}}\,,\qquad
\Phi_{\rm eff} \propto
t^{\frac{\sqrt{25+48Q^2}-5}{6}}\,,
\end{eqnarray}
which grow faster than the solutions
given in Eq.~(\ref{persol1}).
This leads to changes in the matter power spectrum of the
LSS as well as in the ISW effect in the CMB.

The field perturbation $\delta \phi$ is the sum of the
matter-induced mode given in Eq.~(\ref{delphi}) and
the oscillating mode $\delta \phi_{\rm osc}$ given in
Eq.~(\ref{ddotphi}).
Using the WKB solution (\ref{delphiosc}) for
the latter mode, we have
\begin{eqnarray}
\label{delphi2}
\delta \phi=c_1 t^{\frac{\sqrt{25+48Q^2}-5}{6}}
+c_2 t^{-2/3} \cos (c t^{1/3})\,.
\end{eqnarray}
Since the frequency has a dependence
$|\dot{\omega}_\phi/\omega_\phi^2| \simeq H \propto 1/t$,
the WKB approximation tends to be accurate at late times.
Equation (\ref{delphi2}) shows that the matter-induced mode
dominates over the oscillating mode with time.

\subsection{The matter power spectra}

The models (\ref{model}) have a heavy mass $M$
which is much larger than $H$
in the deep matter-dominated epoch, but which gradually
decreases to become of the order of $H$ around the present epoch.
Depending on the modes $k$, the system crosses the point
$M^2/F=k^2/a^2$ at $t=t_k$ during the matter era.
In the context of $f(R)$ gravity this indeed happens for the modes
relevant to the galaxy power spectrum \cite{Tsuji08,TUT}.
Since for the model (\ref{model}) $M$ evolves as 
$M \propto t^{-\frac{2-p}{1-p}}$ during the matter era,
the time $t_k$ has a scale-dependence given by
$t_k \propto k^{-\frac{3(1-p)}{4-p}}$.
When $t<t_k$, the evolution of $\delta_m$ is given by
Eq.~(\ref{persol1}), but for $t>t_k$ its evolution
changes to the form given by~(\ref{persol2}).

We define the growth rate of the matter perturbation to be
\begin{eqnarray}
s \equiv \frac{\dot{\delta}_m}{H \delta_m}\,,
\end{eqnarray}
which is $s=1$ in the regime $M^2/F \gg k^2/a^2$.
After the system enters the regime $M^2/F \ll k^2/a^2$
during the matter-dominated epoch, we have
\begin{eqnarray}
\label{sestimate}
s =\frac{\sqrt{25+48Q^2}-1}{4}\,.
\end{eqnarray}

\begin{figure}
\includegraphics[height=3.6in,width=3.6in]{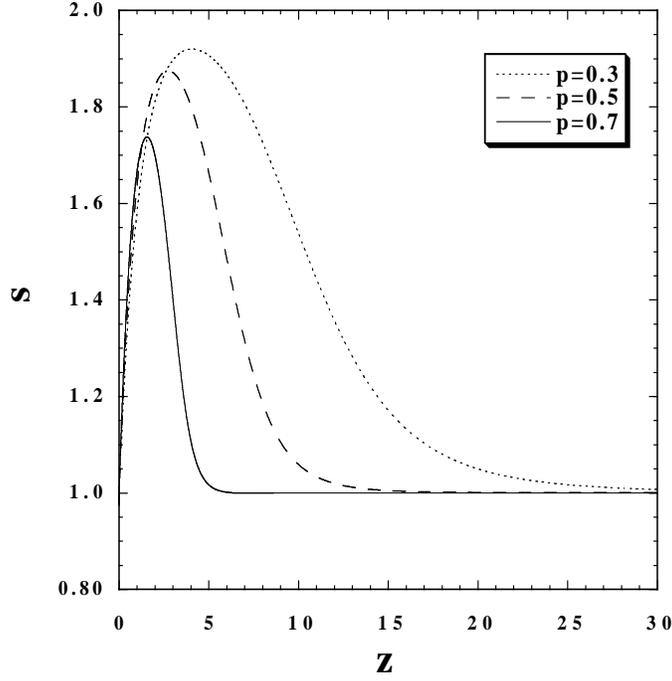}
\caption{
The evolution of the growth rate $s$ of matter perturbations
in terms of the redshift $z$ for $Q=1.08$ and
$k=600a_0H_0$ with three different values of $p$.
For smaller $p$ the critical redshift $z_k$ gets larger.
The growth rate $s$ reaches a maximum value
and begins to decrease after the system enters
the accelerated epoch. For smaller $p$
the maximum value of $s$ tends to approach
the analytic value given in Eq.~(\ref{sestimate}).
}
\label{per}
\end{figure}

During the matter era the mass squared
is approximately given by
\begin{eqnarray}
\label{Mphi2}
M^2 \simeq
\frac{1-p}{(2^p\,p\,C)^{1/(1-p)}}Q^2
\left( \frac{\rho_m}{V_0}
\right)^{\frac{2-p}{1-p}} V_0\,.
\end{eqnarray}
Using the relation $\rho_m=3F_0
\Omega_m^{(0)}H_0^2 (1+z)^3$,
we find that the critical redshift $z_k$ at time
$t_k$ can be estimated as
\begin{eqnarray}
\label{zk}
z_k \simeq \left[\left( \frac{k}{a_0H_0} \frac{1}{Q}
\right)^{2(1-p)} \frac{2^p pC}{(1-p)^{1-p}}
\frac{1}{(3F_0 \Omega_m^{(0)})^{2-p}}
\frac{V_0}{H_0^2} \right]^{\frac{1}{4-p}}-1\,,
\end{eqnarray}
where $a_0$ is the present scale factor.
The critical redshift increases for larger $k/(a_0H_0)$.
The matter power spectrum, in the linear regime,
has been observed for the scales
$0.01h$\,Mpc$^{-1} \lesssim k \lesssim 0.2h$\,
Mpc$^{-1}$, which corresponds to
$30a_0H_0 \lesssim k \lesssim 600a_0H_0$.
In Fig.~\ref{per} we plot the evolution of the
growth rate $s$ for the mode $k=600a_0H_0$
and the coupling $Q=1.08$ with three different values of $p$.
We find that, in these cases, the critical redshift
exists in the region $z_k \gtrsim 1$ and that $z_k$
increases for smaller $p$.
When $p=0.7$ we have, $z_k=3.9$, from
Eq.~(\ref{zk}), which is consistent with the
numerical result in Fig.~\ref{per}.
The growth rate $s$ reaches to a maximum value $s_{\rm max}$
and then begins to decrease around the end of the matter era.

McDonald {\it et al.} \cite{Mc} derived the constraint
$s=1.46 \pm 0.49$ around the redshift, $z=3$, from
the measurement of the matter power spectrum from the
Lyman-$\alpha$ forests.
The more recent data reported by
Viel and Haehnelt \cite{Viel} in the redshift
range $2<z<4$ show that even the value $s=2$ can be
allowed in some of the observations.
The likelihood analysis using these data
for the coupled quintessence scenario gives the
constraint $s \lesssim 1.5$ \cite{Di}.
If we use the criterion $s<2$ for the analytic
estimation (\ref{sestimate}),
we obtain the bound $Q<1.08$.
{}Figure \ref{per} shows that $s_{\rm max}$ is smaller than
the analytic value $s=2$ (which corresponds to $Q=1.08$).
When $p=0.7$, for example, we have that $s_{\rm max}=1.74$.
{}For the values of $p$ that are very close to 1,
$s_{\rm max}$ can be smaller than 1.5.
However these cases are hardly distinguishable from
the $\Lambda$CDM model.
In any case the current observational data on the growth rate $s$
is not enough to place tight bounds on $Q$ and $p$.

The growth of matter perturbations continues to
the time $t_{\Lambda}$ characterized by the
condition $\ddot{a}=0$. At time $t_{\Lambda}$
the matter power spectrum $P_{\delta_m}=(k^3/2\pi^2)
|\delta_m|^2$ shows
a difference compared to the $\Lambda$CDM model given by
\begin{eqnarray}
\frac{P_{\delta_m}(t_\Lambda)}{P_{\delta_m}^{\Lambda {\rm CDM}}}
=\left(\frac{t_{\Lambda}}{t_k}\right)^{2\left( \frac{\sqrt{25+48Q^2}-1}
{6}-\frac23  \right)}
\propto k^{\frac{(1-p)(\sqrt{25+48Q^2}-5)}{4-p}}\,.
\end{eqnarray}
The CMB power spectrum is also affected by the non-standard
evolution of $\Phi_{\rm eff}$ given in Eq.~(\ref{persol2}).
This mainly happens for low multipoles because of the
ISW effect. Since the smaller scale modes in CMB relevant to
the galaxy power spectrum are hardly
affected by this modification, there is a difference between
the spectral indices of the matter power
spectrum and of the CMB spectrum
on the scales, $k>0.01h$\,Mpc$^{-1}$:
\begin{eqnarray}
\label{deln}
\Delta n (t_{\Lambda})=
\frac{(1-p)(\sqrt{25+48Q^2}-5)}{4-p}\,.
\end{eqnarray}
This reproduces the result in the $f(R)$ gravity
derived in Ref.~\cite{Star07}.
In Ref.~\cite{Tsuji08} it was further shown that this
analytic estimation agrees well with numerical results
except for large values of $p$ close to unity.
This reflects the fact that for larger $p$
the redshift $z=z_k$ at time $t=t_k$
gets smaller (being of the order of $z_k={\cal O}(1)$)
so the approximations used in deriving the solution
(\ref{persol2}), based on $w_{\rm eff}=0$ and
$\Omega_m=1$, break down.
In Ref.~\cite{Tsuji08} it was also found that
the difference $\Delta n (t_0)$ integrated to the
present epoch does not show significant
difference compared to (\ref{deln}).

\begin{figure}
\includegraphics[height=3.8in,width=3.8in]{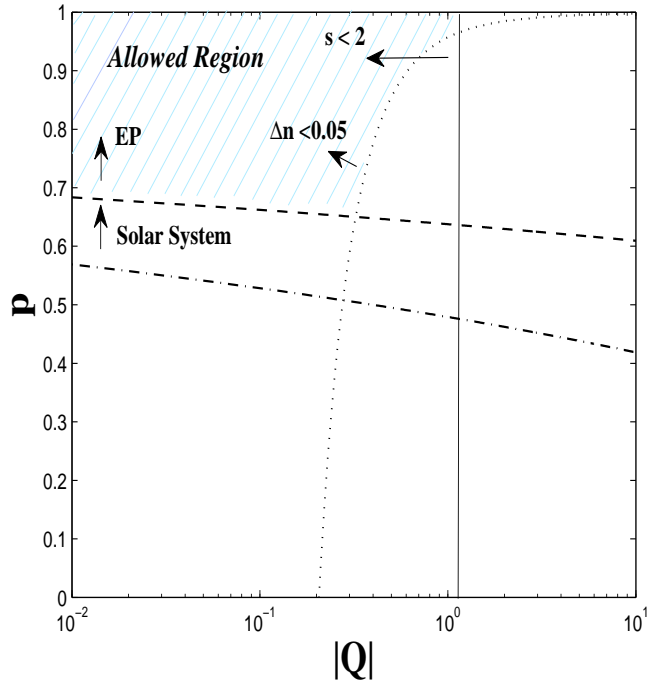}
\caption{
The allowed region of the parameter space in the
$(p, Q)$ plane. We show the bounds coming from
the conditions $\Delta n (t_\Lambda)<0.05$ and
$s<2$ as well as the solar-system constraint
(\ref{solar}) and the EP constraint (\ref{EP}).
}
\label{allowed}
\end{figure}

At present we do not have any observationally
significant evidence for the presence of a
difference between the spectral indices of
the CMB and the matter power spectra \cite{Teg}.
In Fig.~\ref{allowed} we plot the constraints coming from
the criterion, $\Delta n (t_\Lambda)<0.05$.
If $|Q|$ is smaller than 0.1, this condition is
trivially satisfied.
For larger $|Q|$ the constraints
on the values of $p$ tend to be stronger.
In the $f(R)$ gravity we obtain the bound
$p>0.78$, which is stronger than the constraint
coming from the violation of the equivalence principle.
If we adopt the criterion $\Delta n (t_\Lambda)<0.03$,
the bound on $p$ becomes tighter: $p>0.87$.
Meanwhile, if $|Q|$ is smaller than the order of 0.1,
the EP constraint gives the tightest bound.
If we use the criterion $s<2$ for the analytic estimation
(\ref{sestimate}) then the coupling $|Q|$ is bounded from
above ($Q<1.08$).

In Fig.~\ref{allowed} we show the allowed parameter
space consistent with current observational and
experimental constraints.
The constraints coming from the ISW effect in the CMB
due to the change in evolution of the
gravitational potential do not provide
tighter bounds compared
to those shown in Fig.~\ref{allowed}.

\section{Conclusions}
\label{conclude}

We have considered a class of dark energy models
based on scalar-tensor theories given by the
action ~(\ref{action2}).
In these theories, expressed in the Einstein frame,
the scalar field $\phi$ is coupled to
the non-relativistic matter with a constant coupling $Q$.
The action (\ref{action2}) is equivalent to the Brans-Dicke
theory with a field potential $V$, where the Brans-Dicke parameter
$\omega_{\rm BD}$ is related to the coupling
$Q$ via the relation $3+2\omega_{\rm BD}=1/2Q^2$.
These theories include the $f(R)$ gravity theories
and the quintessence models as
special cases where the coupling is given by
$Q=-1/\sqrt{6}$ (i.e., $\omega_{\rm BD}=0$)
and $Q=0$  (i.e., $\omega_{\rm BD} \to \infty$), respectively.

We began by studying
the background cosmological
dynamics in a homogeneous and isotropic setting,
without specifying the field potential $V(\phi)$ but
under the assumption that the slope
of the potential, $\lambda=-V_{,\phi}/V$, is constant.
The varying $\lambda$ case can also be studied
by treating the fixed points as instantaneous ones.
We found that for a range of values of the coupling constant $|Q|$
not much smaller than unity the matter era can be realized by the
solution corresponding to the point (d) in Eq.~(\ref{scaling})
subject to the condition $\lambda/Q \gg 1$.
Interestingly the presence of a non-zero coupling $Q$ leads to
a de-Sitter solution characterized by the condition
$V_{,\phi}+QFR=0$ (i.e., $\lambda=4Q$),
which can lead to late-time acceleration.
(The condition for the stability of this de-Sitter solution
is given by ${\rm d}\lambda/{\rm d}\phi<0$
at the fixed point.)

In the absence of the scalar-field potential,
solar-system tests constrain the coupling $Q$
to have values in the range $|Q| <2.5 \times 10^{-3}$.
The presence of the potential, on the other hand,
allows the LGC to be satisfied for larger values of $|Q|$, if the field is
sufficiently heavy in the high-curvature region where
gravity experiments are carried out.
We found that even when $|Q|$ is of the order of 1,
a thin-shell can form inside a spherically symmetric body
such that the effective coupling $|Q_{\rm eff}|$
defined in Eq.~(\ref{phir1}) becomes much smaller than 1.

We then considered a family of models
given by the scalar-field potentials (\ref{model})
which generalize the corresponding potential in the $f(R)$ theory, while
at the same time satisfying the LGC for appropriate choices of the
parameters. In particular we found that as
$p$ approaches unity, the mass of the field $\phi$ becomes larger,
thus allowing the LGC to be satisfied more easily [see Eq.~(\ref{Mphi})].
Using the constraints coming from solar system tests as well as
compatibility with the equivalence principle, we obtained the
bounds $p>1-5/(9.6-{\rm log}_{10} |Q|)$ and
$p>1-5/(13.8-{\rm log}_{10} |Q|)$, respectively.
In the $f(R)$ gravity, for example, these constraints
correspond to $p>0.50$ and $p>0.65$ respectively.

During radiation/matter eras the field $\phi$
needs to be very close to 0 for the compatibility
with LGC, which results in, $F=e^{-2Q\phi} \simeq 1$.
Figure \ref{allowed} summaries the regions of
the parameter space in the $(p,Q)$ plane where
the corresponding potentials lead to
models compatible with the LGC.

For these models we found that
the quantity $F$ tends to increase from its present value
as we go into the past, which results
in the equation of state $w_{\rm DE}$
of dark energy becoming singular when
$\Omega_m=F_0/F$. This behavior is similar to that
found for $f(R)$ theories.

We also studied the evolution of density perturbations for these
models in order to place constraints on the coupling $Q$ as well as
on the parameters of the field potential.
In the deep matter era the mass $M$ of the scalar field
is sufficiently heavy to make these models compatible with LGC,
but it gradually gets smaller as the Universe enters
the accelerated epoch.
For the models compatible with the galaxy
power spectrum,  there exists a ``General Relativistic'' phase
during the matter era characterized by
the condition $M^2/F \gg k^2/a^2$.
At this stage the matter perturbation $\delta_m$ and
the effective gravitational potential $\Phi_{\rm eff}$
evolve as $\delta_m \propto t^{2/3}$ and
$\Phi_{\rm eff}={\rm constant}$, respectively,
as in the case of Einstein gravity.
Around the end of the matter-dominated epoch,
the deviation from the Einstein gravity can be seen
once $M^2/F$ becomes smaller than $k^2/a^2$.
The evolution of perturbations during this ``scalar-tensor''
regime is given by Eqs.~(\ref{persol2}).
Under the criterion $s=\dot{\delta}_m/H\delta_m<2$
of the growth rate of matter perturbations with the use of
the analytic estimation (\ref{sestimate}),
we obtain the bound $Q<1.08$.
The difference $\Delta n$ of the spectral indices of
the CMB and the matter power spectra gives rise
to another constraint on the model parameter $p$
and the coupling $Q$.

{} Figure \ref{allowed} illustrates the bounds
derived from the conditions $\Delta n<0.05$ and $s<2$,
as well as those from local gravity constraints.
The models with $p$ close to 1 satisfy all these requirements.
It will be certainly of interest to place more stringent
constraints on the values of $p$ and $Q$
by using the recent data of matter power spectrum,
CMB and lyman alpha forest.
Moreover, the future survey of weak lensing
may find some evidence of an anisotropic stress
between gravitational potentials $\Phi$ and $\Psi$,
which can be a powerful tool to distinguish
modified gravity models from the
$\Lambda$CDM cosmology.

\section*{ACKNOWLEDGEMENTS}
K.~U. would like to thank
Gunma National College and The University of Tokyo for 
their kind hospitality during
his stay where a part of this work was done.
J.\ Y.\ would like to thank Bernard J.\ Carr for his
kind hospitality at Queen Mary College where a part of
the work was done.
K.~U. is supported by the
Science and Technology Facilities Council (STFC) and
S.\,M. is supported by a JSPS Research Fellowship.
This work was partially supported by JSPS
Grant-in-Aid for Scientific Research
Nos.~30318802(ST), 16340076(JY), and 19340054(JY).

\appendix
\section{Stability analysis}

In this Appendix, we briefly summarizes the results
of the stability analysis which are necessary
to obtain the background cosmological dynamics
discussed in Sec.~{\ref{sechomcos}}.

\subsection{Stability of the fixed points}
\label{apstab_ana}

When $\lambda$ is a constant,
one can analyze the stability of the critical points (a)-(e)
[i.e., Eqs.~(\ref{fp1})-(\ref{fp5})] by considering small perturbations
$\delta x_1$ and $\delta x_2$
around them \cite{CST}.
We write the equations for perturbations in the form
\begin{eqnarray}
\frac{\rd }{\rd N}
\left(
\begin{array}{c}
\delta x_1 \\
\delta x_2
\end{array}
\right) = {\cal M} \left(
\begin{array}{c}
\delta x_1 \\
\delta x_2
\end{array}
\right) \,,
\label{uvdif}
\end{eqnarray}
and derive eigenvalues $\mu_1$ and $\mu_2$
of the matrix ${\cal M}$ to assess
the stability of fixed points.
They are given by
\begin{itemize}
\item (a)
\begin{eqnarray}
\mu_1=-\frac{3-2Q^2}{2(1-2Q^2)}\,,\quad
\mu_2=\frac{3+2Q\lambda-6Q^2}{2(1-2Q^2)}\,.
\end{eqnarray}
\item (b1)
\begin{eqnarray}
\mu_1=\frac{3(\sqrt{6}+4Q-\lambda)}{\sqrt{6}+6Q}\,,
\quad
\mu_2=\frac{3+\sqrt{6}Q}{1+\sqrt{6}Q}\,.
\end{eqnarray}
\item (b2)
\begin{eqnarray}
\mu_1=\frac{3(\sqrt{6}-4Q+\lambda)}{\sqrt{6}-6Q}\,,
\quad
\mu_2=\frac{3-\sqrt{6}Q}{1-\sqrt{6}Q}\,.
\end{eqnarray}
\item (c)
\begin{eqnarray}
\mu_1=-\frac{6-\lambda^2+8Q\lambda-16Q^2}{2(1-4Q^2+Q\lambda)}\,,
\quad
\mu_2=-\frac{3-\lambda^2+7Q\lambda-12Q^2}{1-4Q^2+Q\lambda}\,.
\end{eqnarray}
\item (d)
\begin{eqnarray}
\mu_{1,2}=\frac{3(2Q-\lambda)}{4\lambda}
\left[ 1\pm \sqrt{1+\frac{8(6Q^2-2Q\lambda-3)
(12Q^2+\lambda^2-7Q\lambda-3)}{3(2Q-\lambda)^2}} \right]\,.
\end{eqnarray}
\item (e)
\begin{eqnarray}
\mu_1=\mu_2=-3\,.
\end{eqnarray}
\end{itemize}
\subsection{Stability of the de-Sitter point for the variable $\lambda$}
\label{des_st}

While the point (e) is stable for constant $\lambda$, it is
not obvious that this property also holds for a varying
$\lambda$. In what follows we shall discuss the stability
of the de-Sitter point.

It is convenient to consider the variable $\lambda (\phi)$
as a function of $F(\phi)$, i.e., $\lambda=\lambda(F)$.
We define a variable, $x_4 \equiv F$, that satisfies the
following equation
\begin{eqnarray}
\label{dx4}
\frac{{\rm d}x_4}{{\rm d}N}=-2\sqrt{6}
Q x_1 x_4\,,
\end{eqnarray}
where the r.h.s. vanishes at the de-Sitter point (e).
Considering the $3 \times 3$ matrix for perturbations
$\delta x_1$, $\delta x_2$ and $\delta x_4$
around the point (e), we obtain the eigenvalues
\begin{eqnarray}
\mu_1=-3\,,\quad
\mu_{2,3}=-\frac32 \left[ 1 \pm
\sqrt{1-\frac83 F_1 Q
\frac{{\rm d} \lambda}{{\rm d} F} (F_1)}
\right]\,,
\end{eqnarray}
where, $F_1\equiv F(\phi_1)$
is the value of $F$ at the de-Sitter point with the field value $\phi_1$.
Since, $F_1>0$, we find that the de-Sitter point
is stable for
\begin{eqnarray}
\label{decon}
Q \frac{{\rm d} \lambda}{{\rm d} F}
(F_1)>0\,,\quad {\rm i.e.}, \quad
\frac{{\rm d}\lambda}{{\rm d} \phi}
(\phi_1)<0\,.
\end{eqnarray}
We checked that this agrees with the stability condition
derived in Refs.~\cite{Faraoni} by considering metric
perturbations about the de-Sitter point.

In $f(R)$ gravity this condition translates into
${\rm d} \lambda/{\rm d}F<0$.
Since in this case, $F=e^{2\phi/\sqrt{6}}={\rm d}f/{\rm d}R$ and
$V=(RF-f)/2$, we have $\lambda=-R f_{,R}/\sqrt{6}V$.
Then, together with the fact that $Rf_{,R}=2f$ holds
for the de-Sitter point, the condition,
${\rm d} \lambda/{\rm d}F<0$,
is equivalent to $R<f_{,R}/f_{,RR}$.
For positive $R$ this gives
\begin{eqnarray}
0<\frac{Rf_{,RR}}{f_{,R}}<1\,,
\end{eqnarray}
which agrees with the stability condition
for the de-Sitter point derived in Ref.~\cite{AGPT}.

\end{document}